%% file: automatic-connectors.tex
\documentclass{vldb}

\PassOptionsToPackage{pdfpagelabels=false}{hyperref}
\usepackage{xspace}
\usepackage{xcolor}
\usepackage{algorithm}
\usepackage[noend]{algpseudocode}
\usepackage{listings}
\usepackage{subfigure}
\usepackage{textcase}
\usepackage{multirow}
\usepackage{booktabs}
\usepackage{array}
\usepackage{hyperref}
\usepackage{inconsolata}
\usepackage{times-lite}
\usepackage{paralist}
\usepackage{flushend}
\usepackage[T1]{fontenc}

\newcommand{\Name}[0]{PipeGen\xspace}
\newcommand{\IORedirect}[0]{IORedirect\xspace}
\newcommand{\FormOpt}[0]{FormOpt\xspace}

\newcommand{\NAME}[0]{\texorpdfstring{\MakeUppercase{\Name\xspace}}\xspace}
\newcommand{\DBMS}[0]{DBMS\xspace}
\newcommand{\DBMSs}[0]{DBMSs\xspace}
\newcommand{\CSV}[0]{CSV\xspace}
\newcommand{\HDFS}[0]{HDFS\xspace}
\newcommand{\IO}[0]{IO\xspace}
\newcommand{\FileOutputStream}[0]{\texttt{FileOutputStream}\xspace}

\newcommand{\URIs}[0]{URIs\xspace}
\newcommand{\AugmentedString}[0]{\texttt{AString}\xspace}
\newcommand{\AugmentedStrings}[0]{\texttt{AString}s\xspace}

\newcommand{\Myria}[0]{Myria DBMS\xspace} %DBMS X (Anonymized)\xspace}

\newcommand{\linelabel}[1]{\label{lst:#1}}

\newcommand{\secref}[1]{Section~\ref{sec:#1}}  % for use in text
\newcommand{\Secref}[1]{Section~\ref{sec:#1}}  % for start of sentence
\newcommand{\figref}[1]{Figure~\ref{fig:#1}}     % for use in text
\newcommand{\Figref}[1]{Figure~\ref{fig:#1}}   % for start of sentence

\newcommand{\lineref}[1]{line \ref{lst:#1}}  % for use in text
\newcommand{\tabref}[1]{Table~\ref{t:#1}}  % for use in text
\newcommand{\eg}{{e.g.}}
\newcommand{\ie}{{i.e.}}

\def\thepage{\arabic{page}} % delete for final camera-ready

\def\sharedaffiliation{%
\end{tabular}
\begin{tabular}{c}}

\begin{document}

\title{\Name: Data Pipe Generator for Hybrid Analytics}
\numberofauthors{3}
\author{
  \alignauthor Brandon Haynes\\
  \alignauthor Alvin Cheung\\
  \alignauthor Magdalena Balazinska
  \sharedaffiliation
  \affaddr{Department of Computer Science \& Engineering} \\
  \affaddr{University of Washington} \\
  \affaddr{\{bhaynes, akcheung, magda\}@cs.washington.edu}
}

\date{1 November 2015}

\maketitle
\begin{sloppypar}
\begin{abstract}

  % Moving data between database systems with different storage formats,
  % capabilities, and query languages has become increasingly common in
  % modern data analytics. A typical way to move data among different systems is to
  % first export the data to be transferred to the file system using a common
  % intermediate format (e.g., \CSV), and then import the same data from the
  % the common intermediate format stored on the file system into the target system.
  % We refer to such data transfer logic as {\em data pipes}. Needless to say,
  % this approach necessitates excessive file
  % system \IO and data conversions. Meanwhile, implementing custom data transfer
  % logic requires repeatedly
  % implementing dedicated data pipes between each pair of
  % systems and is highly non-scalable.

  %In this paper,

  We develop a tool called \Name for efficient data transfer between
  database management systems (\DBMSs). \Name targets data analytics
  workloads on shared-nothing engines. It supports
  scenarios where users seek to perform different parts of an analysis
  in different \DBMSs or want to combine and analyze data stored in
  different systems. The systems may be co-located in the same cluster
  or may be in different clusters. To achieve high performance, \Name
  leverages the ability of all \DBMSs to export, possibly in parallel,
  data into a common data format, such as \CSV or JSON.
  It automatically extends these import and
  export functions with efficient binary data transfer capabilities
  that avoid materializing the transmitted data on the file
  system.  We implement a prototype of \Name and evaluate it by
  automatically generating data pipes between five different
  \DBMSs. Our experiments show that \Name delivers speedups up to
  $3.8\times$ compared with manually exporting and importing data across
  systems using \CSV.

\end{abstract}

%\category{H.2.5}{Database Management}{Heterogeneous Databases}
%\category{I.2.2}{Artificial Intelligence}{Program Transformation, Program Modification}
%\category{D.1.2}{Programming Techniques}{Automatic Programming}
%%Omit this category?\category{D.2.12}{Software Engineering}{Interoperability}[Data mapping]

%\terms{\todo{Performance}}
%\keywords{\todo{Add keywords}}

\section{Introduction}
\label{sec:intro}
\input{introduction}

\section{Motivating Example}
\label{sec:motivation}
\input{motivation}

\section{The \NAME Approach}
\label{sec:approach}
\input{approach}

\section{File IO Redirection}
\label{sec:redirect}
\input{redirection}

\section{Optimizations}
\label{sec:opt}
\input{optimizations}

\section{Implementation}
\label{sec:impl}
\input{implementation}

\section{Evaluation}
\label{sec:eval}
\input{evaluation}

\vspace{-0.3cm}
\section{Related Work}
\label{sec:relwork}
\input{related-work}

\vspace{-0.3cm}
\section{Conclusion}
\label{sec:concl}
\input{conclusion}

\vspace{-0.3cm}
\section*{Acknowledgements}
\vspace{-0.1cm}
\thanks{\small{This work is supported in part by the National Science Foundation through NSF grants IIS-1247469 and IIS-1110370, and gifts from the Intel Science and Technology Center for Big Data, Amazon, and Facebook.}}

\end{sloppypar}

{\scriptsize
  \bibliographystyle{abbrv}
  \vspace{-0.35cm}
  \bibliography{references}
}

\end{document}

%% file: introduction.tex
Modern data analytics often requires retrieving data stored in
multiple database management systems (\DBMSs). As an example, consider
a scientist who collects network data in a graph database such as
Giraph~\cite{avery2011giraph}, while storing other metadata in a
relational database system such as Derby~\cite{derby}. To analyze her
data, the scientist must retrieve data from the relational store, and
subsequently combine it with the data stored in the graph database to
produce the final result.  As part of the process, she might further
need to temporarily move the retrieved data to other \DBMSs to
leverage their specialized capabilities such as machine learning
algorithms~\cite{datoblog}, array processing operators~\cite{scidb},
and so on. This type of federated data analysis is referred to as
\textit{hybrid analytics} and remains poorly supported today even as
the landscape of big data management and analytics systems is rapidly
expanding.

%We refer to such data analytics across multiple systems as
%{\it hybrid analytics}.

%graph analytics~\cite{avery2011giraph}, or

%internal data formats and algorithm implementations, for

In general, a hybrid analytics task involves moving data
among $n$ different \DBMSs, with each \DBMS potentially having its own
internal data format.  A critical requirement of hybrid analytics is
the ability to move data efficiently among the different \DBMSs
involved~\cite{stonebraker:15}.  There have been three main approaches
to solving this problem.  The simplest approach is to use an
intermediate extract transform and load (ETL) process that issues
queries to the source \DBMS using protocols such as JDBC~\cite{JDBC}
or through a specialized wrapper, extracts the data, then inserts it
into the target \DBMS. This approach, however, fails to leverage the
fact that the source and target systems may both be shared-nothing and
are likely executing in the same cluster. It is thus ill-suited for
moving large datasets.

As a faster approach, many \DBMSs provide means for users to export
the stored data, often in parallel, to an external format, and import
the data from that same format. The user can leverage such a capability
to transfer data between the two \DBMSs by first exporting data from
the source system, and subsequently importing it into the target
\DBMS.  This works particularly well in settings where a common data
format exists between the source and target \DBMSs, for instance
text-based data formats such as comma-separated values (\CSV) or
binary data formats such as Protocol Buffers~\cite{protobuf},
Parquet~\cite{parquet}, and Arrow~\cite{arrow}.  Unfortunately, moving
data using a text-oriented format such as CSV or JSON is extremely
costly: it requires serializing the data to be transferred from the
internal format of the source \DBMS to the common text format, storing
the serialized data to physical storage, and finally importing from
storage into the internal format of the target \DBMS.\footnote{In this
  paper, we use \DBMS to refer to both relational and non-relational
  stores. Also, we assume the transferred data will be discarded and
  not persisted in the target \DBMS after the user has examined the
  query results.}  Shared binary formats alleviate some overheads but
still require transforming and materializing the transferred data,
which incurs unnecessary and non-trivial cost in both time and storage
space. Additionally, while support for text-oriented formats such as CSV are
common, shared binary formats remain rare and evolve rapidly. As a
concrete example, CSV is the only common data format supported by the
five \DBMSs (Myria~\cite{myria}, Spark~\cite{spark},
Giraph~\cite{avery2011giraph}, Hadoop~\cite{hadoop}, and
Derby~\cite{derby}) that we use in the evaluation.

%Using physical storage as the intermediary for moving data
%incurs unnecessary and non-trivial cost in both time and storage
%space.

%\magda{This paragraph seems to imply that to build a data type, a user
 % needs to implement JDBC but what about simply using the existing
  %JDBC support and issuing queries to each system that way. We should
 % discuss that scenario here since this is the foundation of federated
 % systems.}

A third approach to moving data that avoids using physical storage as
intermediary is to construct dedicated data transfer programs, \ie,
{\it data pipes}, between specific source and target \DBMSs. This
ranges from implementations of common data transfer protocols such as
JDBC~\cite{JDBC}, ODBC~\cite{ODBC}, and Thrift~\cite{thrift}, to
software packages for transferring data between two specific systems,
such as \texttt{spark-sframe} for moving data between Spark and
GraphLab~\cite{datoblog}.  Unfortunately, generalizing this approach
requires implementing $\mathcal{O}(n^2)$ data pipes in order to
transfer data between $n$ different \DBMSs. In addition to being
impractical, this approach requires knowledge about the internals of
the source and target \DBMSs, making it inaccessible to non-technical
users. Even for technical professionals, implementing dedicated data
pipes is often an error-prone process; the pipe implementations are
often brittle as systems evolve over time and this renders previous
pipe implementations obsolete.

In this paper, we describe a new tool called \Name that retains the
benefits of dedicated data pipes without the shortcomings of manual
implementation or serializing via physical storage.  Using a
combination of static and dynamic code analysis, \Name automatically
constructs and embeds a data pipe into a \DBMS by exploiting
the fact that many \DBMSs can export and import data from
commonly-used formats (e.g., CSV), and that most systems come with
unit tests that exercise that functionality.  \Name takes a number of
inputs, including a set of such unit tests and the source code of the
\DBMS. \Name analyzes the source code of the \DBMS and executes each
of the export unit tests to create a data pipe that redirects data
being exported to the disk to a network socket that is provided
by \Name at runtime.  Similarly, the import unit tests
are used to create a data pipe that reads data from a network socket
rather than a disk file. The same unit tests are used to validate the
correctness of the generated data pipes. Users then leverage the
generated pipes by issuing queries that export and import from disk
files just like before, except that \Name connects the generated pipes
together via sockets so that data can be transmitted directly
from the source to the target \DBMS, bypassing the disk, and
leveraging parallelism when possible.

\Name comes with a number of optimizations to improve the performance
of the generated data pipes. In particular, \Name extends the
text-oriented data format implementations with extra code that
\textit{intercepts the original data} before its text encoding,
identifies and eliminates delimiters, and identifies and eliminates
redundant metadata. By capturing the original data, \Name
enables the transmission of that data using an efficient binary format
(Apache Arrow~\cite{arrow} in our prototype implementation). \Name
thus uses existing CSV or JSON data export and import capabilities as
a specification to automatically implement an efficient, binary data
pipe. With this approach, \Name not only enables efficient binary data
transfers but also simplifies the evolution of the binary data
transfer formats: whenever a new format becomes available
(\eg, the next generation of Apache Arrow), it suffices to update the
\Name tool and that new format automatically becomes available to
transfer data between different big data systems, without having to
implement support for the new format in each of these systems
separately.

%First, \Name attempts to identify and remove
%delimiters from text-oriented data formats to remove unnecessary costs in serializing
%delimiters. In addition, \Name transfers data between \DBMS using a custom
%binary format, and removes extraneous computation (e.g., serializing data types
%to strings) for additional speed up.

Our experiments show that \Name can generate data pipes that speed up
data transfer between \DBMSs by factors up to $3.8\times$, both when
the source and target systems are colocated on a single node, and when
they are placed on separate machines.

\Name currently does not address the problem of schema matching and
focuses on the mechanics of data movement instead. We expect the
generated data pipes to be utilized directly by the end user or a
query optimizer, which would insert additional operators to perform
schema matching either before or after data has been transferred.

In summary, this paper makes the following contributions:
\begin{itemize}

\item We describe an approach based on program analysis
to automatically create data pipes and enable direct data transfer between \DBMSs
(\secref{approach} and~\secref{redirect}).

%\item For both cases where a pair of \DBMSs share only a text-oriented
%external format (e.g., \CSV or \todo{JSON}) or

\item We present a number of techniques that improve
the performance of the generated data pipes by removing
the unnecessary computation that are inherent in text-oriented formats (\Secref{opt}).
Importantly, these optimizations enable \Name to automatically
replace an inefficient text-based data pipe with an efficient, binary one.
Through this approach, \Name makes fast data
transfer formats available to data analytics engines without requiring
developers to manually change the source code of these engines.

\item We build a prototype of \Name, and evaluate it by
  generating data pipes between five different Java-based \DBMSs.  The
  optimized data pipes are up to $3.8\times$ faster than a na\"{i}ve
  approach that transfers data via the file system using CSV.

 % using \CSV and \todo{JSON}
 % as a target for our textually-oriented optimizations.

%Our optimization techniques
 % further improve the performance of the data pipes by up to $30\%$,
 % and the performances of the final optimized pipes are within
 % \todo{$15\%$} of a manually-constructed variant that utilizes
 % Protocol Buffers as an intermediate format (\Secref{impl} and \Secref{eval}).
 % \alvin{why highlight protobufs?}

\end{itemize}

% In the following, we first describe an example that highlights the
% importance of our approach in~\secref{motivation}.  In~\secref{approach}, we introduce \Name
% and give an overview of our approach.  \Secref{redirect}
% and \secref{opt} then describe the steps to generate data
% pipes.  We describe our \Name prototype in~\secref{impl}, evaluate
% its performance in~\secref{eval}, and discuss related work in
% \secref{relwork}.

%% file: motivation.tex
\begin{figure}
\begin{center}
\includegraphics[width=\columnwidth]{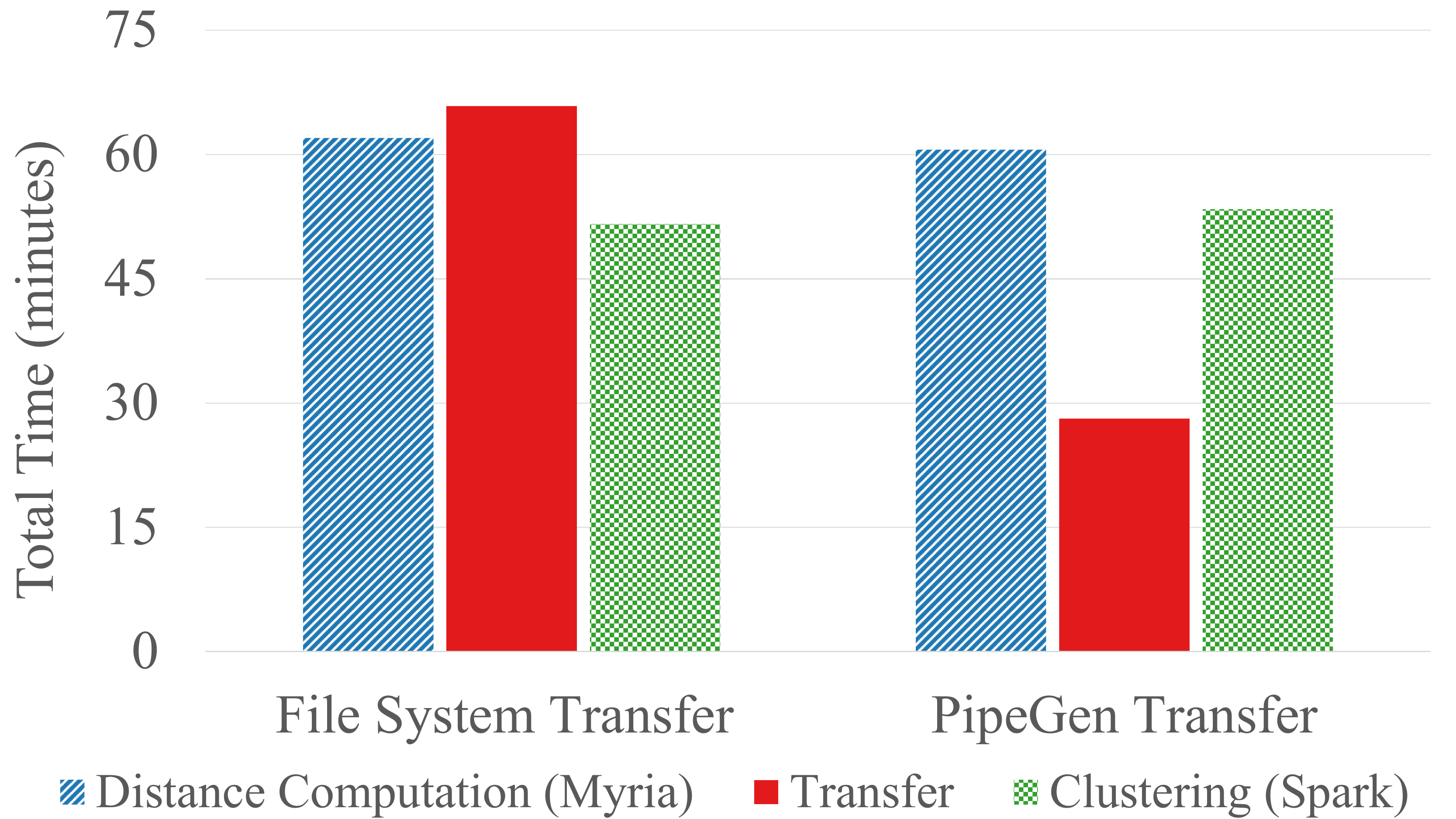}
\end{center}
\vspace{-0.2in}
\caption{Total time for an astronomy workflow with and without
\Name-generated data pipes.  Pairwise distance between each tuple pair was computed
in Myria and transferred to Spark where PIC clustering~\protect\cite{lin2010power} was performed.}
\label{fig:clustering-performance}
\vspace{-0.1in}
\end{figure}

%\magda{We don't need to anonymize for VLDB.}

% In the previous section, we highlighted the use of hybrid analysis for
% scientists who wish to utilize functionalities available on a \DBMS
% external to where their data is stored.  To make this concrete,
% in this section, we present a real-world astronomy workflow that illustrates
% such a hybrid analysis and shows how using data pipes automatically
% generated with \Name yields a performance advantage compared to using
% existing data export and import capabilities.

Consider an astronomer who studies the evolution of galaxies using
N-body simulations of the universe~\cite{jetley2008massively}. A
typical workflow for this analysis is for the astronomer to first
collect the output of a cosmological simulation.  The output takes the
form of a series of snapshots of the universe.  Each snapshot consists
of particles distributed in 3D space. The astronomer then needs to
cluster the particles for each snapshot into galaxies in order to
analyze the evolution of these galaxies over time. In this example,
the particle clustering is a critical piece of the analysis and
astronomers often experiment with different clustering
algorithms~\cite{knebe2013structure}.

In our scenario, an astronomer uses the Myria \DBMS~\cite{myria,myriawebsite}
to analyze the output of
an N-body simulation, including performing an initial particle
clustering, when she learns of a novel clustering method that may be
of interest.  This new method may or may not outperform the current
approach, and the astronomer is interested in evaluating the two.
Critically, however, Myria \emph{does not support this technique},
but it is available in Spark~\cite{spark}.

A typical approach involves the following: perform some or all of the
data preparation tasks in Myria, export the intermediate result to
the file system, and import the files into
Spark.  Since the only common file format supported by both systems
is \CSV (both also support JSON but produce incompatible documents),
the intermediate result is materialized as a set of \CSV files. A second
iteration is necessary to bring results back to Myria for
comparison.

Unfortunately, exporting and importing large data via the file system
is an expensive proposition, and for some datasets there may be
insufficient space available for a second materialization. The
alternate approach of modifying the \DBMS source code to support
efficient data transfer requires deep programming and database systems
expertise.  Unfortunately, this data movement challenge makes it
very difficult or impossible for domain scientists to leverage the
highly-specialized functionalities across different \DBMSs.

The \Name tool avoids these disadvantages.  To illustrate, we execute
the workflow described above on the present-day snapshot of a 5TB
astronomy simulation stored in
the Myria. The present day snapshot
is 100~GB in size and an initial data clustering has already been
performed on the data. Our goal is to count differences in cluster
assignments between the existing data clustering and those for power
iteration clustering (PIC)~\cite{lin2010power} available on
Spark.  We compute pairwise distances in
Myria between all particles having distance less than threshold
$\epsilon=0.00024$, set by our astronomy collaborators.  This yields
approximately one billion pairs.  We then transfer this data to Spark
where we perform PIC clustering.  Finally, we transfer the
resulting assignments back to Myria for comparison with the
existing clusters.  To contrast different data transfer mechanisms,
we perform the transfer using both the file system as intermediary
and data pipes generated by \Name.

\Figref{clustering-performance} illustrates the performance of each of
these steps.  When using the file system, transfer of the large set of
intermediate pairwise differences is more expensive than either of the other
two steps alone.  Using the data pipe generated by \Name, however,
reduces the time spent in data transfer from 66 to 28 minutes. This result
suggests the importance of having efficient data transfers in support of hybrid
data analytics, and we explain how \Name generates
data pipes automatically in the following sections.

%% file: approach.tex
\Name enables users to move data efficiently between \DBMSs.
%improved performance or to take advantage of functionality on a remote \DBMS.
In this section, we give an overview of \Name's usage and approach.

%\magda{Reminder: discuss somewhere the case where engine A exports to
 % CSV while engine B imports from JSON. We should also say what
 % happens if engine A exports to both CSV and JSON.}

\subsection{Using \Name}

%\Name has an offline phase and an online phase. In the offline phase,
%\Name takes as input the source code of a \DBMS and extends that
%source code with an efficient data pipe to other engines.  In the
%online phase, a user leverages these newly-generated pipes for more
%efficient cross-\DBMS query execution.

To use data pipes, a user first invokes \Name to generate the source code
for data pipes. After compilation, users can write queries to leverage
the generated pipes. Below we give an overview of the two steps.

\noindent
\textbf{Constructing a data pipe.} To generate a data pipe,
the user invokes \Name with a number of inputs. These include
(1) a script used to build the \DBMS along with its source code, (2) a pair of
scripts that execute unit tests associated with import and export
functionality for a specific data format (\eg, CSV, JSON, or binary format),
(3) the name
of the specific data format produced during import and export (for use during
library substitution; see~\secref{library-extension}),
(4) additional configuration-related metadata.

Given these inputs, \Name analyzes the source code of the \DBMS, and generates
a data pipe that exports data via to the network socket to be instantiated by \Name
during runtime.

%  is a series of modifications to
%the source code of the \DBMS that enable data pipe usage (see
%\secrefs{redirect}{opt}) in the online phase.

\begin{figure}[t]
\centering
\includegraphics[width=\columnwidth]{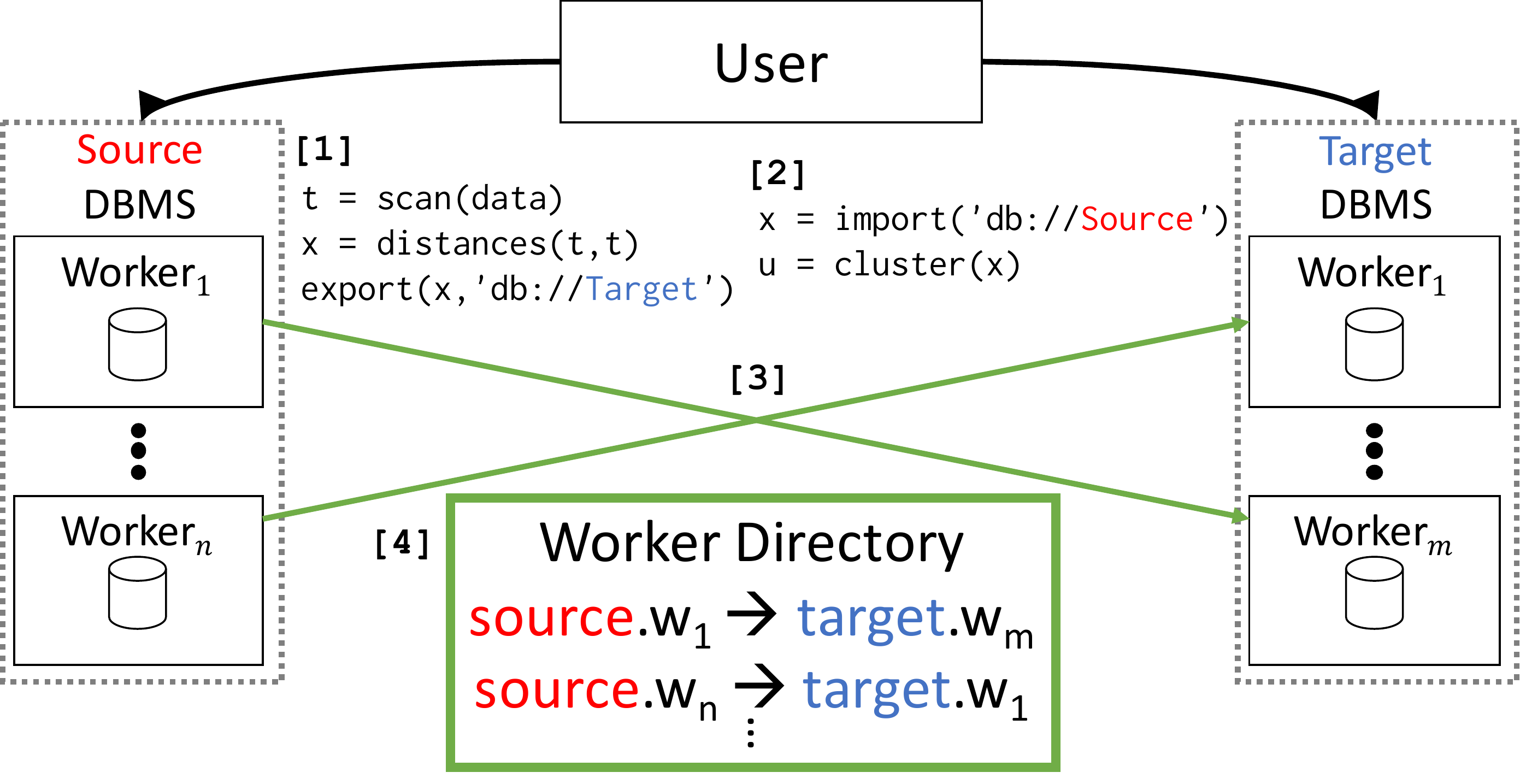}
\vspace{-0.75cm}
\caption{Using the data pipe generated by \Name for the
hybrid analysis from~\secref{motivation}: 1. User submits a query to the
source \DBMS (\eg, Myria) to compute distance and export to the target \DBMS using data pipe.
2. User issues an import query on the target \DBMS (\eg,
Spark) to cluster the result.
Data is transferred using the generated data pipe in 3.,
and in 4. a worker directory (see~\secref{worker-directory})
coordinates the connection process. \Name-related components are highlighted in green.}
\label{fig:datapipe-workflow}
\end{figure}

\noindent
\textbf{Using the generated data pipe.}
To use the data pipe generated by \Name, the user
executes two queries: one that exports data from the source \DBMS, and one that
imports data into the target \DBMS, as shown in
\figref{datapipe-workflow}.
These queries may occur in any order; \Name
will automatically block until both \DBMSs are
ready (see~\secref{worker-directory}).
Note that the queries issued to each \DBMS are
written as if data are moved from the source to target \DBMS via physical storage:
users only need to specify the data to be exported from the source and the name of
the target \DBMS using the special ``\texttt{db://X}'' syntax rather than a filename.

After receiving the queries, \Name connects the generated data pipes
in the source and target \DBMSs by passing them a network socket to transmit data.
In the case where the source and target \DBMSs support multiple worker threads,
they are matched with each other by the worker directory maintained by
\Name, as shown in~\figref{datapipe-workflow}.

\textbf{Usage in the context of hybrid and federated systems.} Because of the
flexibility that \Name's approach affords, we also expect that our
automatically-generated data pipes are well-suited for use in other
contexts such as by being integrated into an existing hybrid (e.g.,
\cite{elmore2015demonstration, cyclops}) or federated \DBMS (e.g.,
Garlic~\cite{josifovski2002garlic}).

\begin{figure}[t]
\centering
\includegraphics[width=\columnwidth]{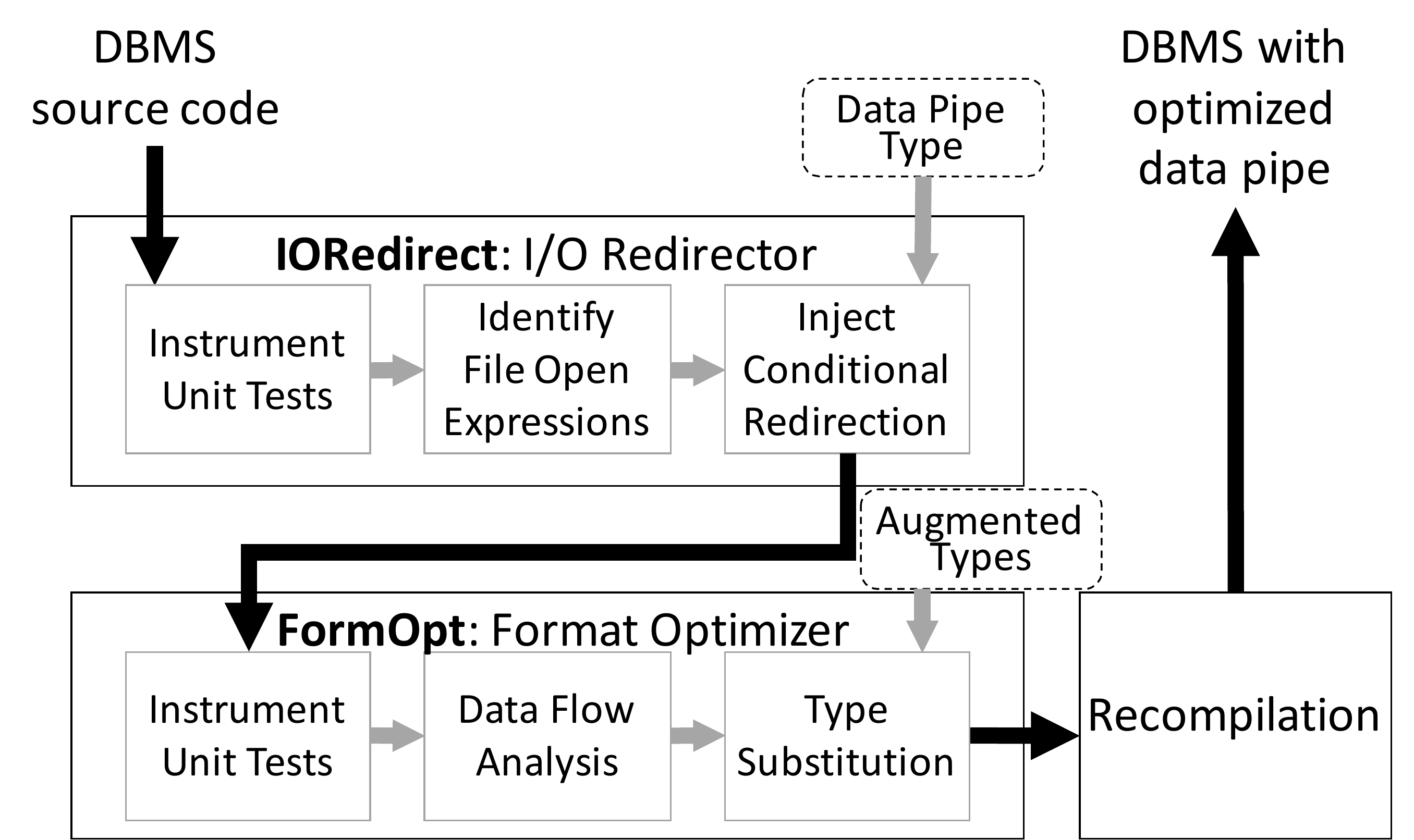}
\vspace{-0.6cm}
\caption{Compile-time components of \Name.  The file IO
  redirector (\IORedirect)
generates a data pipe to transfer data via a network socket, while the format
optimizer (\FormOpt) improves the efficiency of data transfer for
text-oriented formats.}
% \todo{Replace ``Augmented DBMS'' with ``DBMS with Optimized Data Pipe''} \todo{Label boxes with correct names (Connection Redirector -> \IORedirect, Format Optimizer -> \FormOpt)}}
\label{fig:workflow}
\end{figure}

\subsection{Data Pipe Generation Overview}
\label{sec:overview}
To generate a data pipe, \Name takes as
input the source code of a \DBMS and tests that exercise the import and export
functionality
for a given format. It begins by executing and analyzing the provided tests.
Using the results, it modifies the source code of the \DBMS to generate a
data pipe that can transfer data directly between the \DBMSs using a common
data format. When the common format is
text-oriented, \Name further optimizes the format of the transferred
data to improve performance, as we will discuss in~\secref{opt}.

%These steps take place at compile time and are
%designed to be transparently included in the build process of the
%input \DBMSs. \brandon{@Alvin -- per your comments in Figure 3: should I remove this sentence too?}

\Name supports single-node and parallel \DBMSs. For the latter, as illustrated in
\figref{datapipe-workflow}, the source code produced transfers data in parallel directly
between individual workers. Our implementation and evaluation
currently target shared-nothing systems, but the approach can be
applied to other parallel architectures.

To generate a data pipe, \Name performs
the two-phase compilation process shown in~\figref{workflow}.
First, \Name locates in the \DBMS's source code the relevant code fragments
that import from and export data to disk files.
For each of these fragments, \Name identifies the file
system operations (\eg, file open and
close) and substitutes them with equivalents on a network socket. It
accomplishes this by instrumenting the provided unit tests and
identifying the expressions in the code that open the import or export file on disk.
It then modifies these expressions to allow direct sending and receiving
of data via a network socket instead of going through the file system.
This approach enables datasets of arbitrary size to be transmitted (in
contrast to an approach based on memory-mapped files, for example),
and for file systems that do not support named pipes (e.g., \HDFS).
To further improve the performance of the generated data pipe, the
format optimizer eliminates unnecessary
operations for text-oriented formats.
We respectively explain both phases in detail in the following sections.

\noindent
\textbf{Assumptions.}
%\Name targets existing logic in a \DBMS that reads from and writes to
%the file system in a particular format. Because of this,
\Name assumes that the implementation of a given \DBMS's data import and
export is well-behaved.
For instance, \Name requires that the implementation not perform
large seeks during file IO and that an import or export file is opened exactly
once, since otherwise multiple sockets will be created during pipe generation.
For text-based formats, \Name assumes that character
strings are used to serialize data (rather than
byte arrays, for instance), and that values are constructed through
string concatenation rather than random access.
\Name also assumes that, for a given transfer, each \DBMS support the same data format.
Finally, \Name assumes that the unit tests fully exercise the code associated with
the import and export of data.  For export, this includes all code paths between reading data in the
internal representation and
serializing them to disk using system calls.  For import this includes reading serialized data from disk
back into the internal data representation.

%% file: redirection.tex
In this section, we discuss the redirection component of \Name.
%, which serves to produce a basic data pipe.
This component, called
\textit{IORedirect}, creates a data pipe from the \DBMS's existing serialization
code and modifies that code such that instead of using
the file system, data
are exported to (and imported from) a network socket provided by \Name
at runtime.
%The generated source code automatically creates the
%socket during query execution and,
When the source and
target \DBMSs are colocated on the same machine, the socket is
simply a local loopback one. Otherwise the socket connects to the
remote machine hosting the target \DBMS, with the address provided by
a directory (\secref{worker-directory}).

%replaces the file system operations (e.g., reads
%and writes) in the \DBMS source code that are relevant for data import
%and export with equivalents that operate on sockets
%instead. \autoref{fig:modified-architecture} illustrates the
%transformation.

\subsection{Basic Operations}

\label{sec:redirect:single}

%IORedirect accomplishes this replacement by

%To do so, it leverages the DBMS's existing capability to
%import/export data from the file system.  Each \DBMS with such
%capability contains calls in its source code that opens, reads or
%writes, and closes the underlying data file.

In order to modify a DBMS's source code in this way,
%From the serialization code from the \DBMS's source code,
\IORedirect first identifies the
relevant file system call sites that open the disk file used for
import or export. It does not suffice
to search the source code for ``open file'' operations,
however, because a \DBMS may open multiple, unrelated files such
as debugging logs.
%Since this debug log will not contain data that
%needs to be transmitted,
\IORedirect should
not inadvertently replace such files with sockets.

%In identifying the file open call sites in the \DBMS source code,
%the file IO redirector disambiguates relevant import/export file
%open operations from other unrelated occurrences.

\begin{figure}[t]
\centering
\includegraphics[width=\columnwidth]{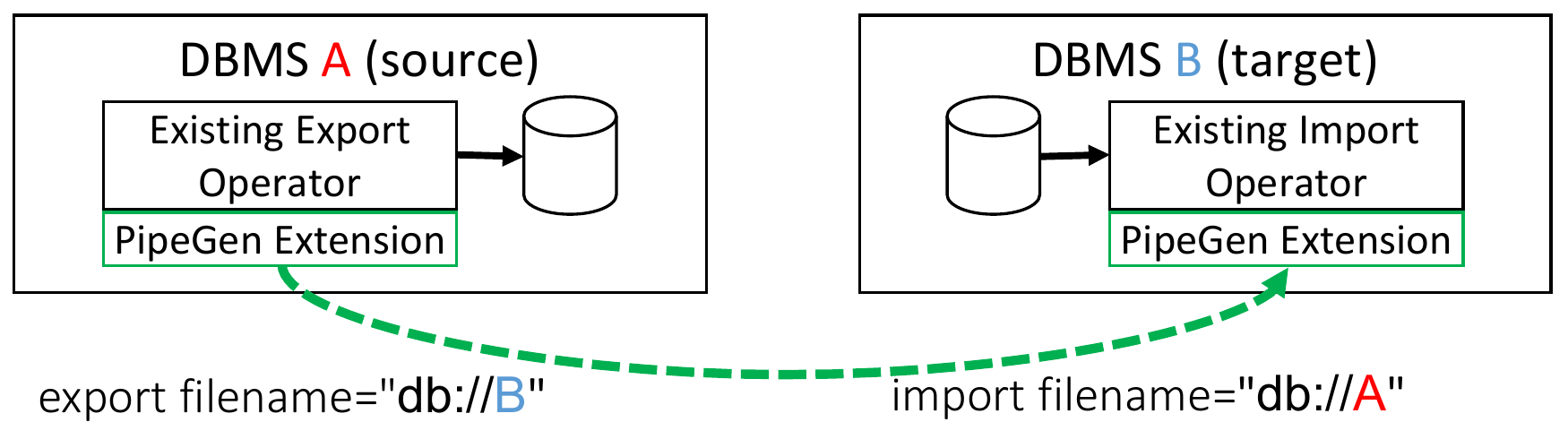}
\vspace{-0.35in}
\caption{\Name creates a data pipe to transmit data via
a network socket. User activates the data pipe by specifying a reserved
filename as the export and import destination.
}
%\todo{Remove ``CSV'' from
 % operator boxes.  Remove worker designations (just ``DBMS x'').}
%\magda{Show all modifications in red or blue. The modifications are:
 % (1) extended export/import operators and (2) the pipe with the data
 % flowing on it.}}
\label{fig:modified-architecture}
\end{figure}

\IORedirect disambiguates this by instrumenting all file open calls in the
source code,
executing the import and export unit tests, and capturing the filename
that is passed in to each call. All calls with filenames
other than the target of the import / export are eliminated.
\IORedirect then modifies each remaining call site by adding code that
redirects I/O operations to a network socket.  The new code is executed
only when a user specifies a \textit{reserved filename} for import or
export.  The conditional nature of the redirection is important
because we want to preserve \DBMS's ability to import from and export to disk files.
%
%\magda{I don't understand the following paragraph. I thought
%redirection used the file names in the unit tests to identify the
%relevant call sites.}
%As an additional defense against redirecting irrelevant files, \Name
%uses a reserved filename format that is unlikely to be activated for
%existing irrelevant call sites.  As a default we use a URI format with
%the custom scheme \texttt{dbms}, and this may be specified by the user
%at runtime to be arbitrarily distinct.
This augmented architecture is
illustrated in~\figref{modified-architecture}.

%Once found, IORedirect modifies these call sites to (optionally)
%switch to sending/receiving the file over a network socket rather than
%via a physical file on the file system.  The switch is important
%because we want to preserve the ability to perform an ordinary, file
%system-oriented import or export.

When adding the conditional logic to each relevant call site,
%To make these call site modifications,
\IORedirect introduces a specialized
data pipe class into the \DBMS's source code that
reads or writes to the remote \DBMS
via a passed-in network socket rather than the file system,
but is otherwise a
language-specific subtype of its file system-oriented counterpart.
This allows \Name to substitute all instances of the file IO classes
with a data pipe instance.  To illustrate,
a Berkeley socket descriptor can be
substituted for a file descriptor in a C or C$++$-like language,
while Java or .NET file streams can be interchanged with network
socket streams.

% Moved to above.
%To use the generated data pipes, a user specifies a reserved filename
%as the import/export filename in the query plan.  For example, as we
%saw in \figref{datapipe-workflow}, when a query requires moving data
%from \DBMS $A$ to \DBMS $B$, the user executes an export query on
%\DBMS $A$ to the reserved filename ``\texttt{dbms://B}'', and a
%corresponding import query on \DBMS $B$ to import from the reserved
%filename ``\texttt{dbms://A}''.  All other filenames continue to
%import/export to the file system as usual.

\IORedirect modifies the code to use the generated
data pipe when
the reserved filename is specified.  \Figref{code-modifications}
illustrates one such modification made by \IORedirect for the
Myria DBMS, which is written in Java.  Here an instantiation of
\FileOutputStream is replaced with a ternary operator that examines
the filename and substitutes a data pipe class (itself a subtype of
\FileOutputStream) upon detecting a match.  Due to subtyping,
other IO operations (e.g., \texttt{close}) operate as before.

\begin{figure}[t]
\small
\begin{center}
%[language=Java, numbers=left, xleftmargin=2em]
\begin{lstlisting}[language=Java]
void export(String filename, ...) {
  ...
  stream = new FileOutputStream(filename);
  ...
  stream.close();
}
\end{lstlisting}
\vspace{-0.1in}
\scalebox{1.5}{$\Downarrow$}
\begin{lstlisting}[language=Java]
void export(String filename, ...) {
  ...
  stream = filename.matches(format)
    ? new DataPipeOutputStream(filename);
    : new FileOutputStream(filename);
  ...
  stream.close();
}
\end{lstlisting}
\vspace{-0.35in}
\end{center}
\caption{An example modification to create a Java data pipe.
Here the unit tests indicate line 3 (top) to be modified.
The modified code below,
matches against the reserved filename format (bottom, line 3), and
uses the generated data pipe if a match is found (bottom, line 4).
Due to subtyping, subsequent uses of the resulting stream are unaffected (e.g., line 7).
}
\label{fig:code-modifications}
\end{figure}

% \alvin{I felt that the paragraph below is redundant.}
%The result of the modifications made by the file IO redirector is
%a \DBMS augmented with a basic data pipe.  When activated with a reserved
%filename,
%the data pipe is able to send/receive data to/from a remote \DBMS.

\Name verifies the correctness of the generated data pipe as follows.  First,
it launches a \textit{verification proxy} that imports and exports data as if
it were a remote \DBMS.  This proxy redirects all data received
over a data pipe to the file system, and transmits data from the file system
through a data pipe.  Next, the unit tests for the modified \DBMS are executed
using a reserved filename format that activates the data pipe for all files.
As data are imported and exported over the data pipe, the proxy
reads and writes to disk as if the data pipe did not exist.  We then rely on
existing unit test logic to verify that the contents that are read from or
written to disk are correct.
%This process is illustrated in \figref{verification-proxy}.

%\begin{figure}
%\centering
%\todo{Verification Proxy}
%\caption{\todo{Caption}}
%\label{fig:verification-proxy}
%\end{figure}

%Note that when a test fails, \Name may incrementally disable optimizations.
%For example, one or more of the text-oriented optimizations described in
%\secref{opt} fail, they may be disabled.

%\Name verifies the soundness of the resulting generated
%data pipe using the following
%procedure.  First, it takes the existing unit tests of the \DBMS that
%exercises data import/export functionality. For data export tests,
%instead of writing to the original file as in the test case, we
%modify the test case by asking it to
%write to the reserved filename instead. This exercises the
%data pipe generated by \Name. Our test infrastructure then runs the
%modified test case, collects
%the data written to the socket, and runs the original test case that exports
%data to the file system. The data generated in both cases are compared
%to ensure correctness. Similar tests are performed on data imports as well.

%\if 0
%adds the filenames used in the unit tests of the \DBMS to
%its set of reserved filenames and associates those names with a degenerate
%destination \DBMS.  This degenerate \DBMS serves as a proxy for the actual file
%system so that when the modified \DBMS connects to it with a data pipe the
%intermediate values are written to disk.  \Name then executes the complete
%suite of unit tests associated with the \DBMS.
%\fi

Finally, \IORedirect exposes a dynamic debugging mode that
probabilistically verifies the correctness
of the generated data pipe at runtime, and may be used when the source and
destination \DBMSs are colocated.  Under this mode, the first $n$ elements
of the transmitted data are both written to the disk using
existing serialization logic, and transmitted over the data pipe.
The receiving \DBMS then reads the data
from the file system and compares it to the values that have been transmitted
over the pipe.  Any deviations trigger a query failure.
%and a user may tune the
%number of elements compared on a per-query basis.

%\subsection{Worker Directory}
%\subsection{Supporting Multi-Node \DBMS}
\subsection{Parallel Data Pipes}
\label{sec:worker-directory}

Many multi-node \DBMSs support importing or exporting data in parallel
using multiple worker threads.
\Name uses a \textit{worker directory} to match up the worker threads in
the source and target \DBMSs. The directory is instantiated by \Name
when the \DBMS starts and is accessible by all \DBMS worker processes.
%In such systems, our file IO redirection logic is
%not yet sufficient to connect sets of workers from two \DBMSs.  The
%reason for this is that an importing worker does not know the location
%of its corresponding exporting worker, and hence cannot connect to it
%using the mechanism described in the previous section.
%The address and
%port for the directory service is drawn from configuration in a well-known
%location on the local file system and we assume that it
%is launched on a node accessible by all \DBMS worker processes.

\begin{figure}
\centering
\includegraphics[width=\columnwidth]{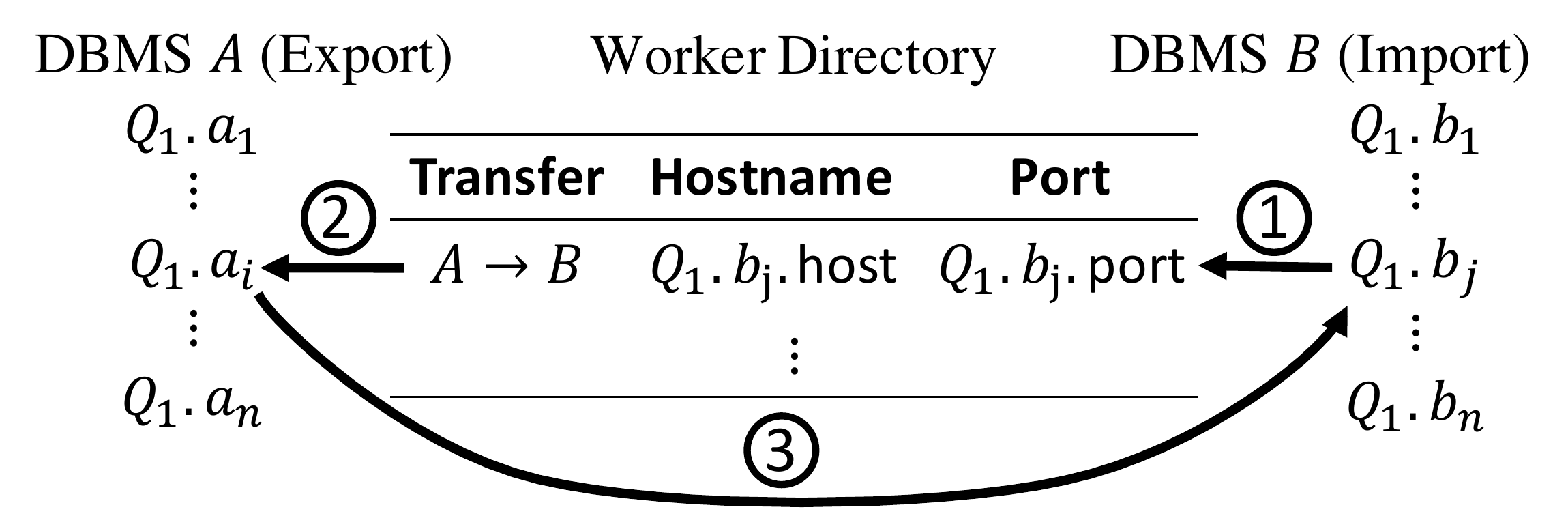}
\vspace{-0.8cm}
\caption{Using the worker directory to initiate data transfer.
In 1., import worker $b_j$ from query $Q_1$ registers with
the directory and awaits a connection.  In 2., export worker $a_i$
queries the directory and blocks until an entry is available,
after which it uses the associated details to connect in 3.}
\label{fig:directory-lookup}
\end{figure}

The worker directory is used as follows.  When a user exports data from
\DBMS $A$ to \DBMS $B$, as \DBMS $B$
prepares to import, each worker $b_1, \ldots, b_n$ registers with the directory
to receive data from \DBMS $A$.
As data is exported from
\DBMS $A$, workers $a_1, \ldots, a_n$ query the directory to obtain the address
and port of a receiving
worker $b_i$.  Each exporting worker $a_i$ blocks until an entry is available
in the directory, after which it connects to its corresponding worker using a
network socket.  To allow multiple concurrent transfers between
the same pair of \DBMSs, each import and export query is also assigned a
unique identifier to disambiguate queries executing simultaneously.
The data pipe class performs the directory registration or query process.
For example, a simplified Java implementation for the query process
is shown in~\figref{directory-code}, and
\figref{directory-lookup} describes the overall workflow.

\begin{figure}
\centering
\begin{lstlisting}[language=Java]
class DataPipeOutStream extends FileOutputStream {
  DataPipeOutputStream(String fname) {
    e = Directory.query(fname); @\linelabel{query}@
    socket = new Socket(e.hostname, e.port); @\linelabel{socket}@
    ...
} }
\end{lstlisting}
\vspace{-0.5cm}
\caption{Simplified Java implementation of directory querying logic.
The query (\lineref{query}) blocks until an
import worker has registered its details with the directory.  The
resulting directory entry is used to open a socket to the remote DBMS (\lineref{socket}).}
\label{fig:directory-code}
\end{figure}

The worker directory ordinarily assumes that the number of workers
between the source and target \DBMSs are identical.  However, a user
may embed metadata to explicitly indicate the number of exporting and
importing processes.  For example, an export
from \DBMS $A$ using two workers to \DBMS $B$ with three workers would use the
respective reserved filenames \texttt{db://A?workers=2} and
\texttt{db://A?workers=3}.
When the number of importing workers exceeds the exporters, the worker
directory opens
a ``stub'' socket for the orphaned importing worker that immediately signals an
end-of-file condition; under this approach the extra importing workers simply sit idle
until the data transfer has completed.
The case where the number of exporting workers exceeds
the importing workers is left as a future extension.

%% file: optimizations.tex
%The algorithm described in the previous section allows \Name to
%utilize the existing data import/export logic in a \DBMS to
%generate a basic data pipe that avoids going through the disk.

When a pair of \DBMSs share an optimized binary format, the data
pipe that \IORedirect generates can immediately be used to
transfer data between the two systems.  For example, \Name can generate
a data pipe that transfers data between Spark and
Giraph using Hadoop sequence files, as that format is supported by both systems.
Unfortunately,
%due in part to the rapid evolution of binary
%serialization formats,
it is rare for two systems to
support the same efficient binary format.  For example, the
Apache Parquet format is (natively) supported by only two of the
systems we evaluate (Spark~\cite{spark} and Hadoop~\cite{hadoop}).
 %Support for modern formats in less frequently updated \DBMSs is even
%less common. For example,
Giraph does not support Parquet without
third-party extension while Apache Derby~\cite{derby} does not
support import and export of any binary format.

By contrast, support for text-oriented formats is far more pervasive:
\textit{all} of the \DBMSs we evaluate (Myria~\cite{myria}, Spark
\cite{spark}, Giraph~\cite{avery2011giraph}, Hadoop
\cite{hadoop}, and Derby~\cite{derby}) support bulk
import and export using a \CSV format, and all but Derby support JSON.
Accordingly,
%when a transfer is
%initiated between a pair of systems that shares no efficient binary
%format,
we expect users to frequently revert to text-based data formats to transfer
data
between such systems.
%will be forced to fall back to a
%more common but less efficient text-oriented format such as these.

Unfortunately, text-based
formats are inefficient and incur
substantial performance overhead.  The main sources of this include:

\begin{itemize}

\item \textbf{String encoding of numeric types.}
It is often the case that the size of a converted string greatly
exceeds that of its original value, which increases transfer time.
For example, the 4-byte floating point value
$-2.2250738585072020 \cdot 10^{-308}$ requires 24 bytes to encode as a string.

\item \textbf{String conversion and parsing overhead.}  Text-oriented export
and import logic spends a substantial amount of time
converting primitive values into/from their string representation.

\item \textbf{Extraneous delimiters.}  For attributes with fixed-width
representations, many data formats include extraneous separators such as value delimiters and newlines.

\item \textbf{Row-orientation.}  \DBMSs that output \CSV and JSON often do so
in a row-oriented manner; this is the case for all of the \DBMSs we evaluate
in \secref{eval}.  This makes it difficult to apply layout and
compression techniques that benefit from a column-oriented layout. For instance,
our experiments show that converting intermediate data to column-major form offers a modest performance benefit (see \figref{library-extension}).

%For data
%pipes, none of this is needed as data can be directly transmitted using
%binary format, as in the case of custom serializers~\cite{protocolBuffers}.

%\alvin{what if the two systems use different encodings? e.g.,
%big vs small endian.  \brandon{I don't explicitly address that in the implementation since I'm all Java.  We could push encoding/endianness into the URI metadata and translate accordingly, but I didn't do any of that.  I suppose I should comment on that in the implementation section?}}

\end{itemize}

%Since \CSV is widely supported among \DBMSs, \Name currently optimizes
%for this format.
%We assume that internal binary import/export formats (which are less frequently
%supported between pairs of \DBMSs) are efficient for data transfer
%purposes.

To improve the performance of text-based
formats, \Name comes with a format optimizer called \textit{\FormOpt}
(see~\figref{workflow}), to address the inefficiencies listed above.
%to the source code of the \DBMS after the
%modifications made by the \IORedirect component. \FormOpt addresses
%all four inefficiencies above as we describe in the rest of this section.
% This component, called \textit{\FormOpt}, optimizes the data pipe
%implementation in three ways: it (1) alters the intermediate format to
%transmit an efficient binary representation of values instead of their string
%representation (\secrefs{string-decoration}{library-extension}), (2) eliminates string conversion overhead for
%primitive types (\secrefs{string-decoration}{library-extension}), (3) removes
%extraneous delimiters and redundant metadata
%(\secref{optimization-intermediate-format}), and (4) applies compression, along with optionally converting intermediate data to compacted column-major form (\secr%ef{optimization-compression}).
\FormOpt optimizes a given text format in one of two
modes. First, it analyzes the \DBMS source code to
determine if the export or import logic use an external
library to serialize data to JSON or \CSV (\Name currently supports
the Jackson JSON library~\cite{jackson} and we plan to support
  Java JSONArrays and the Apache Commons library~\cite{apache-commons-csv}).
If so, then \FormOpt
replaces use of the library with a PipeGen-aware variant.  The implementation
of this
variant is then directly responsible for performing the optimizations
described above.

On the other hand, a \DBMS might
directly implement serialization functionality instead of using an external
library.  For example, the Myria \DBMS directly implements its JSON export
functionality.
%By way of example, for the JSON format Spark uses the Jackson JSON
%processor~\cite{jackson} and Giraph leverages \texttt{JsonArray}s; similarly,
%for CSV Myria uses the Apache CSV parser.
For such systems,
\FormOpt supports a second mode that
leverages \textit{string decoration} to target the actual string
production and consumption logic that takes place during data export and
import. Since components of the string decoration mode are used when applying
the library extension strategy,
we introduce it first and then turn to the library extension process.

%that attempts to
%detect whether a supported serialization library is utilized during
%the process of generating or consuming intermediate textual data.  If
%\FormOpt identifies use of a supported library, it

\subsection{String Decorations}
\label{sec:string-decoration}

%\subsection{Fixed-Width Primitives}
%\label{sec:opt1}

Without using external libraries, we assume the \DBMS uses the following
string operations to serialize data into \CSV or JSON text:
convert objects and primitive values into
strings, concatenate the strings, intersperse delimiters and metadata such as
attribute names, and write the result out. To optimize these steps,
\FormOpt modifies the \DBMS
source code such that, whenever the modified \DBMS attempts to write
the string representation of a fixed-width primitive to the stream, it
instead writes to a compact binary
representation provided by \Name.
Ordinary strings and other non-primitive
values are transmitted in their unmodified form.
As we will see,
doing so eliminates the transmission of unnecessary values such as
delimiters and attribute names.
%Similarly, \FormOpt
%eliminates the transmission of delimiters and removes common metadata such as
%attribute names that are repeated for every row (see
%\secref{optimization-intermediate-format}).

%To avoid the overhead of transmitting strings, \FormOpt substitutes the
%compact binary representation of each fixed-width primitive whenever
%the modified \DBMS attempts to write its equivalent string
%representation.

To accomplish this source code modification, \Name's data pipe type
(introduced in \secref{redirect:single}) must receive the primitive values
of the data to transmit before they are converted into strings.
The core difficulty in ensuring this property is
that there may be many primitive values embedded in any given string
that is passed to the data pipe type.  For example, a particular \DBMS
might concatenate together all attributes before writing them to the
output stream {\tt s}: \texttt{s.write(attr1.toString() + }``\texttt{,}''\texttt{ +
  attr2\-.toString() + ...)}.  By the time the data pipe type
receives the concatenated value, it will be too late as both of the
attributes would have already been converted into strings.

\FormOpt addresses this by introducing a new augmented string called \AugmentedString,
which is a subtype of {\tt java.lang.String}.
\AugmentedString is backed
by an array of {\tt Object}s rather than characters.
By substituting {\tt java.lang.String} instances for \AugmentedStrings in
the right locations, \FormOpt avoids the problem described above by storing
references to the objects to be serialized rather than their string representation.

For example, given ordinary string concatenation:
\vspace{-0.3cm}
\begin{center}
\begin{tabular}{c}
\begin{small}
{\begin{lstlisting}[language=Java, numbers=none]
s = new Integer(1).toString() + "," + "a";
\end{lstlisting}}
\end{small}
\end{tabular}
\end{center}

\FormOpt changes the statement into one that uses \AugmentedStrings:
\vspace{-0.3cm}
\begin{center}
\begin{tabular}{c}
\begin{small}
{\begin{lstlisting}[language=Java, numbers=none]
s = new AString(1) + new AString(",") +
    new AString("a");
\end{lstlisting}}
\end{small}
\end{tabular}
\end{center}

Each of the three instances maintains its associated value as an
internal field ({\tt 1}, {\tt ","}, and {\tt "a"} respectively) and
the concatenated result\textemdash itself an \AugmentedString
instance\textemdash internally maintains the state \texttt{[1, ",",
  "a"]}.  Note that the final \AugmentedString instance need not
include the concatenated string ``1,a'' in its internal state since it
may easily reproduce (and memoize) it on demand.
%\magda{Does
%the following statement mean that more complex types get translated
%into strings? } \AugmentedString
%must also implicitly convert non-primitive values to their ordinary
%string representation so that future changes to their state do not
%impact the version in our array.
More complex types are immediately converted into strings during this
aggregation process to ensure that subsequent changes to their state
do not affect the internal state of the \AugmentedString instance.  However,
as we shall see below, converting a complex object into a string (\eg, through
a \texttt{toString} invocation) may produce an \AugmentedString instance, which
allows for nesting under supported formats.

When an IO method is invoked on the data pipe type, its implementation
inspects any string parameter to see if it is an \AugmentedString.
Additionally, methods exposed by the data pipe type that produce a
string return an \AugmentedString.  During export, this allows the data pipe type to
directly utilize the unconverted values present in the internal state of an
\AugmentedString; similarly, during import the \AugmentedString
implementation
efficiently executes common operations such as splitting on a delimiter and
conversion to numeric values without materializing as character string.

This resolves
the problem we described above, but does not address the issue of
\emph{where} to substitute an \AugmentedString for a regular string
instance.  Intuitively, we want to substitute only values that are
(directly or indirectly) related to data pipe operations, rather than
replacing all string instances in the source code.  To find this
subset, \Name executes each of the provided unit tests and marks all call
sites where data is written to/read from the data pipe.
%(the relevant file IO call sites).
\Name then
performs data-flow analysis to identify the sources of those values
(for export) and conversions to primitive values (for import).  This
produces a data-flow graph (DFG) that identifies candidate expressions
for substitutions.

\begin{figure*}[t]
\begin{center}
\center
\includegraphics[width=0.75\textwidth]{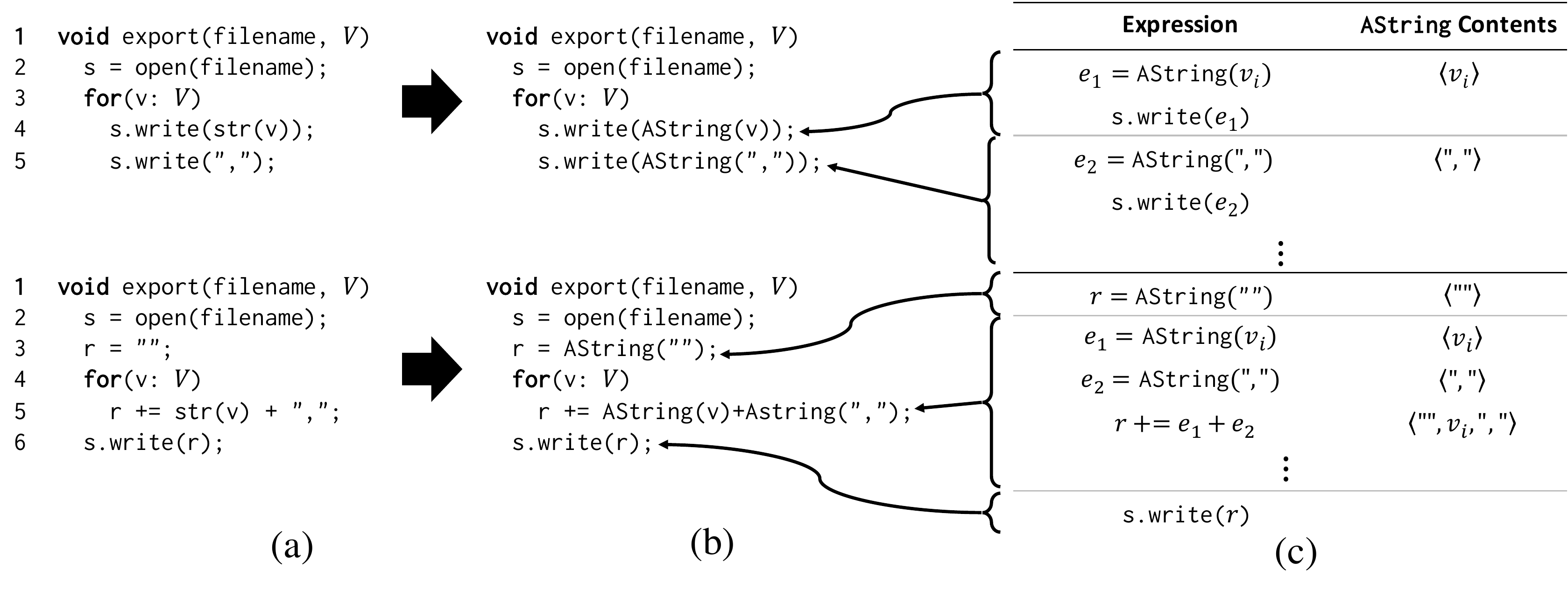}
\end{center}
\vspace{-0.2in}
\caption{(a) Two ways to implement \CSV data export
(inserting newlines instead of comma on the last
value is omitted for clarity); (b) after string replacement of
literals and conversions;
%Each use of \AugmentedString is in the DFG of the stream write
%invocation;
(c) accumulated values in the \AugmentedString instances after one loop
iteration.}
%\todo{Change to actual code (will it fit?)}\\
%\todo{No longer using $S$ as the augmented string type}
\label{fig:vector-accumulation}
\end{figure*}

Using the resulting DFG, \FormOpt replaces three types of
string expressions: string literals and constructors,
conversion/coercion of any value to a string (for export), and
conversion/coercion of any string to a primitive (for import).
%  For example, if
%a literal comma appeared in the DFG, the format optimizer would substitute
%an instance of our augmented type as \texttt{new
%AugmentedString(",")}.
To illustrate this,
\figref{vector-accumulation} shows (a) two
potential implementations of an export function, (b) the code after
string replacement, and (c) the accumulated values in the internal state of
the \AugmentedStrings after one iteration of the loop.

Algorithm~\ref{alg:transform} summarizes the replacement process.  On
lines 1-2, the algorithm executes each test and identifies the relevant file
IO call
sites.  On line 4, it uses these sites to construct a DFG.
For each expression $e$ that converts to or from a string format, it
replaces $e$ with a corresponding \AugmentedString operation (lines
5-14).
Lines 6-9 target expressions relevant for data export; for example, a string
literal $v$ is replaced with an augmented
instance \texttt{AString($v$)}.  To support efficient imports, the
algorithm performs a similar
replacement for strings converted to primitive values (lines 10-14).

%\begin{algorithm}[endfloat]
\begin{algorithm}[t]
\caption{String decoration}
\label{alg:transform}
\scriptsize
\begin{algorithmic}[1]
\Statex{\hspace{-.15in}{\bf function} \textsc{transform}($T$: tests)}
\For {\textbf{each} $t \in T$}
  \State Find relevant IO call sites $C$
  \For {\textbf{each} $c \in C$}
    \State Construct data-flow graph $G$
    \For {\textbf{each} expression $e \in G$}
      \If {$e$ is literal $v$ or\\
           \hspace{1.85cm}$e$ is instantiation \texttt{String}($v$) or\\
           \hspace{1.85cm}$e$ is $v$\texttt{.toString()}}
           \State Replace $v$ with \texttt{AString($v$)}
      \ElsIf{$e$ is \texttt{Integer.parseInt($v$)}}
           \State Replace $v$ with \texttt{AString.parseInt($v$)}
      \ElsIf{$e$ is \texttt{Float.parseFloat($v$)}}
           \State Replace $v$ with \texttt{AString.parseFloat($v$)}
      \EndIf
      \State Similarly for other string operations
    \EndFor
  \EndFor
\EndFor
\end{algorithmic}
\end{algorithm}

As in before, \Name verifies the correctness of the
modifications described above by executing the specified unit tests in
conjunction with the verification proxy. \Name turns off this optimization if
one or more unit
tests fail following the modifications made by the \FormOpt component in
string decoration mode (we have not encountered such cases in our experiments).

\subsection{Using External Libraries}
\label{sec:library-extension}

As mentioned, many \DBMSs use external libraries to
serialize data. Handling such engines require a custom subtype
to be implemented for each external library (of which there are a few that are
commonly used).
%Since there are generally a modest number of popular
%libraries for conversion to a given format, it is reasonable to target
%our optimizations at this level.  Doing so allows \FormOpt to capture
%data at a much higher level than with its string decoration mode and
%modify fewer lines of code.  However, this approach requires that a
%custom subtype be implemented for each supported library.
%
%\magda{Ideally, we should compare the two modes in the
 % evaluation and indicate here the benefit of targeting the latter
 % over the former based on the evaluation results. \brandon{One benefit is
  %fewer lines of code modified.  Table 1 should support this, even if I can't
  %establish a performance difference.}}
%
Under this mode, \FormOpt replaces instantiations of a given formatting
library with a PipeGen-aware subtype that tries to avoid
the overhead associated with strings and delimiters.
For example, whenever the
\DBMS invokes a method that builds or parses the text
format, \Name instead internally constructs or produces a binary
representation.  When the resulting text fragment is converted to string form,
the PipeGen-aware subtype generates an \AugmentedString that contains
the binary representation as its internal state.  During import, the library
subtype recognizes that it is interacting with the data pipe type (q.v.
\autoref{sec:redirect:single}) and directly consumes the intermediate binary
representation.  This allows the library subtype to construct an efficient
internal representation of the input.

%Just as with string decoration, under library substitution the data
%pipe type (q.v. \autoref{sec:redirect:single}) must receive a binary
%representation of the underlying values prior to their conversion to
%text form.  \magda{So how do we address the fact that it must receive
%  a binary representation? It doesn't say.} When a library offers a method to
%convert
%a newly-constructed document to string form during export, this is
%straightforward -- we arrange to return an \AugmentedString instead of
%an ordinary string instance, and when this augmented version is
%eventually exported we leverage the internal state to transmit values
%efficiently.  During import, we perform a similar operation during
%document parsing to emulate the internal state of the library we are
%substituting.  When a library offers other mechanisms to produce or
%consume serialized data (e.g., via a writer or stream interface), we
%support that by introducing a pipe-aware augmented version of that
%mechanism and decorating the involved instances.

As before, \FormOpt must identify only those
locations where a library is used for import and export.  Our approach for doing
so -- using unit tests and DFGs --
is similar to that of string decoration.
%  First, we instrument
%the provided unit tests and identify call sites where data is written to/read
%from the data pipe.  We then produce a DFG over these statements and identify
%expressions where a library is instantiated.  \FormOpt then looks for
%instantiations associated with the set of libraries supported for a given
%format.
For example, if a user specifies JSON when invoking
\Name, \FormOpt will examine the source code of the \DBMS for instantiations
of supported JSON libraries.
Using the resulting DFG, \FormOpt replaces string literals, constructors, and
conversion/coercion expressions in a manner identical to that discussed
above.  Additionally, \FormOpt replaces the instantiation of the library
itself with an augmented variant along with any writer or stream interfaces
that the library exposes.  For example, consider the following simplified
version of the Spark JSON export function:
\vspace{-0.5cm}
\begin{center}
\begin{tabular}{c}
\begin{scriptsize}
{\begin{lstlisting}[language=Java, numbers=none]
String toJSON(RDD<String> rdd) {
  Writer w = new CharArrayWriter();
  JsonGenerator g = new JsonGenerator(w);
  foreach(Object e: rdd) { generateJSON(g, e); }
  return w.toString(); }
\end{lstlisting}}
\end{scriptsize}
\end{tabular}
\end{center}

Our transformations produce the following modified variant:
\vspace{-0.5cm}
\begin{center}
\begin{tabular}{c}
\begin{scriptsize}
{\begin{lstlisting}[language=Java, numbers=none]
String toJSON(RDD<String> rdd) {
  Writer w = new AWriter(new CharArrayWriter());
  JsonGenerator g =
    new AJsonGenerator(new JsonGenerator(w));
  foreach(Object e: rdd) { generateJSON(g, e); }
  return w.toString(); } // an AString!
\end{lstlisting}}
\end{scriptsize}
\end{tabular}
\end{center}

As in string decoration, \FormOpt disables library call replacement if the
generated code does not pass all unit test cases. If string decoration also
fails to pass the tests, then \Name only generates the basic data pipe as
discussed in~\secref{redirect}.

%\magda{Actually, I wonder if all we need is simply the set of
%  substitutions: CharArrayWrite $\rightarrow$
%  AugmentedWriter. Internally, it is up to the developer to implement
%  whatever optimizations he/she wants.}
%\alvin{I guess the question is what does AJsonGenerator do?}
%\brandon{Intercepting the generator allows us to avoid string parsing/
%conversion costs}
%  Additional libraries may be
%incorporated into the library extension mode of the \FormOpt
%component.  New libraries must expose a builder-pattern interface for
%generation of a document for the given format, and must expose a
%visitor-style interface for consuming an imported document.  \alvin{Is it necessary to mention all this? \brandon{Magda suggested we note what is necessary to add a new library.
%Maybe a good candidate to cut for space.}}

\subsection{Intermediate Format}
\label{sec:optimization-intermediate-format}

Using \AugmentedStrings resolves the first two sources
of overhead introduced above (encoding of numeric types and
conversion/parsing overhead).
In this section, we introduce optimizations
designed to eliminate delimiters and avoid redundant metadata. These
optimizations are implemented inside \Name's data pipe type.

% Both the string decoration and library extension modes exposed by \FormOpt are
% designed to connect unconverted primitive values to the data pipe type.  The
% mechanism used to do so involves embedding the original values into the
% internal state of \AugmentedString instances.  These \AugmentedString
% instances are then passed to the implementation of the data pipe type during
% export, and produced by the data pipe type during import.

% While this resolves the first two sources of overhead introduced above
% (encoding of numeric types and conversion/parsing overhead), it does not
% address the removal of delimiters or the format of data transmitted from
% source to destination system.  In this section we introduce optimizations
% designed to eliminate delimiters and avoid redundant metadata.

\subsubsection{Delimiter Inference and Removal}
\label{sec:opt2}

%\magda{Is this on Figure 3? Is this done by \FormOpt specifically as
 % opposed to PipeGen generally? \brandon{I've clarified that this is handled
 % by the data pipe implementation itself, which is "kind of" part of the
 % FormOpt component :)  It depends on the implementation receiving
 % AugmentedString instances, so it seems appropriate to talk about in this
 % section.}}

%The second optimization is to prevent the data pipe from writing unnecessary
%delimiters.  While the \CSV format explicitly uses commas as delimiters,
%many \DBMSs use other data formats with different delimiters.
%Further, some systems allow the delimiter to be varied on a
%per-query basis (e.g., Derby \cite{derby}).  So, we need to infer
%the delimiter used (if any) for each result transmitted through our data
%pipes.

Text-oriented formats such as \CSV and JSON include delimiters that separate
attributes and denote the start and end of composite types.  In some cases
these delimiters are fixed in advance; for example, square brackets are
used in JSON to indicate an array.  However,
default delimiters often vary on a per-system basis.  This is common under \CSV,
where some systems default to a non-comma delimiter (\eg, Hadoop
uses tab separation by default) or allow the delimiter to be specified by the
user (\eg, Derby).  To eliminate delimiters, \FormOpt needs to first infer
them.
%This requires that the data pipe type, for
%some text formats,
%infers  a compile-time inference process for identifying
%these delimiters.  Once the default delimiters for a \DBMS have been
%identified, \Name requires that all subsequent data pipe transfers use these
%inferred values.
\FormOpt does so by first running the provided unit tests.
During the execution of each test, \FormOpt counts
the length-one strings within the
array and identifies which character is most likely to be the
delimiter.  For example, the array
$[1, \text{"|"}, \text{"a,b"}, \text{"\textbackslash n"}]$ contains exactly
one length-one string (\text{"|"}), and \FormOpt concludes that this is most
likely to be the delimiter.
The input $[1, \text{"|"}, \text{"a"}, \text{"\textbackslash n"}]$ is
ambiguous, since both \text{"|"} and \text{"a"} appear with equal frequency.
In this case, \FormOpt applies, in order, the following tie-breaking
heuristics: (i) prefer non-alphanumeric delimiters, and (ii) prefer earlier
delimiters.  Under both heuristics, \FormOpt would select \text{"|"} as the
final delimiter.

Note that should \FormOpt infer an incorrect
delimiter, invalid data will be transmitted to the remote \DBMS.  In the
previous example, if \FormOpt's selection of \text{"|"} was invalid and the
character \text{"a"} was actually the correct delimiter, it would incorrectly
transmit the tuple $(1, \text{"a"})$ instead of the correct value
$(\text{"1|"}, \text{""})$.  More importantly, this is likely to cause the
unit tests to fail as discussed
in~\secref{redirect}.  This results in \FormOpt disabling the
optimization until the unit tests were extended to fully disambiguate the
inference.

\subsubsection{Redundant Metadata Removal}

More complex text formats such as JSON may not require the delimiter inference
described above, but instead serialize complex composite types such as arrays and
dictionaries.  When producing/consuming JSON or a similar textual format, the
composite types produced by a \DBMS often contain values that are highly
redundant.  For example, consider the following document produced by the Spark \texttt{toJSON} method:
\vspace{-0.1in}
\begin{center}
\begin{tabular}{c}
\begin{scriptsize}
{\begin{lstlisting}[basicstyle=\ttfamily, numbers=none]
{"column1": 1, "column2": "value1"}
{"column1": 2, "column2": "value2"}
{"column1": 3, "column2": "value3"}
\end{lstlisting}}
\end{scriptsize}
\end{tabular}
\end{center}
When such JSON documents are moved between systems, the repeated column names
greatly increase the size of the intermediate transfer.  To
avoid this overhead, \FormOpt modifies the format of the intermediate data to
transmit exactly once the set of keys associated with an array of dictionaries.
In the above example, \FormOpt would transmit the column names
$[\text{"column1"}, \text{"column2"}]$ as a \textit{key header}, and then the
values $[(1, \text{"value1"}), ...]$ as a sequence of pairs.  When importing,
\FormOpt reverses this process to produce the original JSON document(s).

The logic for this transformation is embedded into the JSON state machine
(a subcomponent of the data pipe type) that
is used to consume the \AugmentedString array.  When \FormOpt transitions into
the key state for the first dictionary in an array, it accumulates that key in
the key header.  Once the dictionary has been fully examined, \Name transmits
the key header to the remote \DBMS.  Subsequent dictionaries in that array are
transmitted without keys, so long as they identical to the initial dictionary.
While this approach may be extended to nested JSON documents, our
prototype currently only optimizes top-level dictionaries.

If a new key is encountered in some subsequent dictionary after the key header
has been transmitted, \FormOpt adopts one of two strategies.  First, if the
keys from the new dictionary are a superset of those found in the
key header, \FormOpt appends the new key to the
existing key header.  This addresses the common case where the set of exported
keys was not complete due to, for example, a missing value in the initial
exported dictionary.

A second case occurs when the keys associated with a new dictionary are
disjoint from those in the key header.  This might occur during export from a
schema-free \DBMS, where exported elements have widely varying formats.  In
this case, \FormOpt disables the optimization for the current dictionary and
does not remove keys during its transmission.

\subsection{Column Orientation \& Compression}
\label{sec:optimization-compression}

\DBMSs that output text-oriented formats generally do so in a
row-oriented manner.  For example, a Spark RDD containing $n$ elements
that is exported to \CSV or JSON will generate $n$ lines, each
containing one element in the RDD.  This is also true in the other
systems we evaluate, for both JSON and \CSV formats.
However, once \FormOpt produces an efficient data representation,
we no longer need to transmit data
in row-major form.  For example, the data pipe type can accumulate
blocks of exported data and
transform it to column-major form to improve transfer performance.
Indeed, recent work on column-oriented \DBMSs
suggests that some of the benefits (\eg,
compacting/compression, improved IO
efficiency)~\cite{stonebraker2005c} may also improve performance for
data transfer between \DBMSs.

After examining several formats for the
wire representation of our data (see
\secref{intermediate-format}) we settled on Apache Arrow as the data structure
we transmit, since it performs the best.  To maximize performance, our prototype
accumulates blocks of rows in memory, pivots them into column-major form by
embedding them into a set of Arrow buffers, and transmits these buffers to
the destination \DBMS.  The receiving \DBMS reverses this process.

%% file: implementation.tex
We have implemented a prototype of \Name in Java.
The generated data pipes currently support
\DBMSs that make use of the local file system or \HDFS for data import
and export.

\subsection{File IO Redirection}

The file IO redirector in the current prototype
targets \texttt{FileInputStream} and
\texttt{FileOutputStream} as the relevant file system calls
to be modified. In addition to our
concrete implementation of the data pipe class
(\texttt{DataPipeInput}/\texttt{OutputStream}), we created augmented versions
of the following Java classes:

\begin{asparaitem}
\item \texttt{StringBuilder}/\texttt{Buffer}.  We introduced an array analogous
to the one found in \AugmentedString.

\item \texttt{Output}/\texttt{InputStreamWriter}.  These classes were modified to interact with the \texttt{DataInput}/\texttt{OutputStream} classes and contained overloads for string IO.

\item \texttt{BufferedOutput}/\texttt{InputStream}.  These classes were
augmented to detect when the underlying stream was a data pipe class, and
if so buffering was omitted.

\item \texttt{org.apache.hadoop.io.Text}.  We augmented this class with an
object array in the same way as the \AugmentedString class. We needed
to do so because the Hadoop-based systems uses this class similar
to the ordinary Java strings.

\item \texttt{org.apache.hadoop.hdfs.DFSInput}/\texttt{OutputStream}.  We
produced \HDFS-specific data pipe classes for these classes.  Implementation was
analogous to that of \texttt{DataPipeInput/OutputStream} classes.

\item \texttt{java.sql.ResultSet}.  We replaced the \texttt{getString} methods
with a version aware of our \AugmentedString class.

\end{asparaitem}

Our implementation of \texttt{DataPipeInputStream} supports seeks within one data row.
This is needed to support \HDFS; when the \HDFS client opens a
file, it performs a small read/rewind in order to determine whether the file
being read is a Hadoop sequence file.

Two systems that we evaluated disallowed \URIs with a custom scheme (e.g.,
\texttt{dbms://A}) as an import/export target (Derby and Myria).
For these systems, our prototype supports the ability to specify the reserved
filename template in a configuration file.  In each case we substituted an
alternative of the form ``\texttt{\textbackslash
tmp\textbackslash \_\_reserved\_\_[Name]}''.
These same systems perform an explicit check for the existence of the reserved
filename prior to proceeding with an import.  Similarly, \DBMSs that import
from \HDFS calculate split points by examining files before beginning the
import process.  To protect against failures in each of these cases, \Name
automatically creates stub instances of the reserved filename on the file
system during startup.

\subsection{Augmented String Implementation}

We previously defined \AugmentedString to be a subclass of {\tt java.lang.String},
which is declared to be \texttt{final}.  We work around this issue by replacing
the standard String class with a non-{\tt final} version via dynamic code loading.
%To work around this issue,
%\Name replaces \texttt{java.lang.String} with a cloned version in the source
%code of the \DBMS
%being modified.  This cloned version omits the \texttt{final} modifier.  \Name
%then adjusts the Java bootstrap class loader to utilize the cloned \texttt{java.lang.String} at runtime.

For performance reasons the implementation of \AugmentedString in Java
maintains its array of values as a flat byte array.  These byte
arrays are preallocated on startup and managed internally by \AugmentedString.  For
operations that may not be performed on the raw array (e.g., substring), \AugmentedString
falls back to a materialized string representation.

Finally, since Java does not support operator overloading, \Name
rewrites all string concatenation in the source code using functional forms.
For instance,
for {\tt a+b} where {\tt a} and {\tt b} are both {\tt java.lang.String},
\Name rewrites the expression into
\texttt{new AString(a).\text{concat}(new AString(b))} instead.

%% file: evaluation.tex
We evaluate \Name using benchmarks that show data transfer
times between five Java \DBMSs: Myria~\cite{myria}, Spark 1.5.2~\cite{spark},
Giraph 1.0.0~\cite{avery2011giraph}, Hadoop 2.7.1~\cite{hadoop}, and
Derby 10.12.1.1~\cite{derby}.
The first set of experiments examines the overall performance differences
between the data pipes generated by \Name and importing/exporting
data through the file system
(\autoref{sec:paired-transfer}).  Next, we analyze
the performance gains from each of our optimizations
(\autoref{sec:eval:opt}). We then evaluate the impact of different
data formats transferred between \DBMSs
(\autoref{sec:intermediate-format}), along with the compression method
used during transport (\autoref{sec:compression}).  Finally, we examine the
number of modifications done by each step during data pipe compilation
(\autoref{sec:eval:loc}).

%, and compares performance against a
%\Name-generated data pipes and a manually-constructed, hand-optimized data
%pipe.

Unless otherwise specified, experiments utilize a 16-node cluster of
\texttt{m4.2xlarge} instances in
the Amazon Elastic Compute Cloud. Each node has 4 virtual CPUs,
16 GiB of RAM, and an attached 1TB standard elastic block storage device.  We
deploy the most recent stable release of each \DBMS running under OpenJDK 1.8.
Except for Derby, which is a single-node \DBMS, we deploy each system across
all 16 instances using YARN~\cite{yarn}.  For each pair of \DBMSs, we
colocate workers on each node and assign each YARN container two cores and 8
GiB of RAM.

With the exception of \figref{datatype-speedup},
the experiments in this section all involve the transfer
of $n$ elements with a schema having a unique integer key
in the range $[0,n]$ followed by three $(\text{integer} \in [0, n],
\text{double})$ pairs.
Each 8-byte double was sampled from a standard normal distribution.  For
Giraph, we interpreted the values as a graph having $n$ weighted vertices each
with three random directed edges weighted by the following double.

\subsection{Paired Transfer}
\label{sec:paired-transfer}

We first evaluate the overall benefit of our approach for different
combinations of systems.  \Figref{microbenchmarks} shows the total
transfer time between a source and destination \DBMS using (i) an
export/import through the file system based on functionality provided
by the original \DBMS, and (ii) an export/import using the
\Name-generated data pipes.  For this experiment, we transfer
$10^9$ elements, fix the number of workers/tasks at 16 for each
\DBMS, and enable all optimizations.
Since \CSV is the only common format supported by all
\DBMSs, file system transfers use this format and the \Name data pipes are
generated from \CSV export and import code.

\input{results/microbenchmarks}

As the results show, data pipes significantly
outperform their file system-oriented counterparts.  For this transfer
size, the average speedup over all \DBMSs is $3.2\times$, with
maximum speedup up to $3.8\times$.  This speedup is approximately the
same across all transfer sizes and pairs of \DBMSs.  As shown in
\tabref{data-types}, this speedup is also similar across various cluster sizes.

\input{results/data-types}

This result emphasizes the impact that \Name can have on hybrid data
analytics: without writing a single line of code, a user can get
access to 20 optimized data pipes and speed up data transfers between
any combination of the five systems tested by $3.2\times$ on average.
\Name produces this benefit automatically without requiring that
individual system developers agree on an efficient common data
format. Each system developer only needs to implement CSV export and
import methods.

\begin{figure}[t]
\centering
\includegraphics[width=\columnwidth]{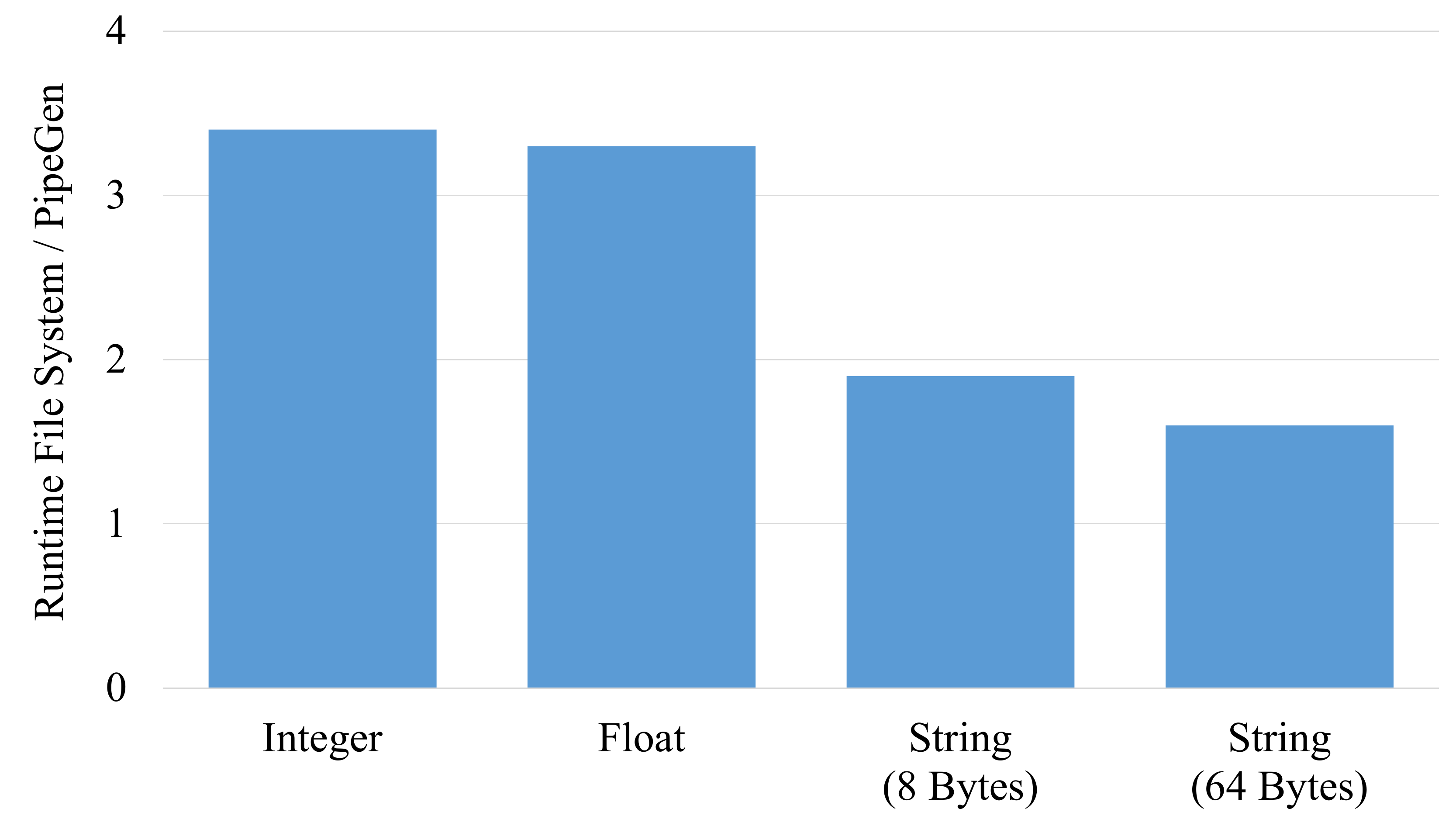}
\vspace{-0.65cm}
\caption{Overall speedup between file system and \Name transfer for different
data types and sizes.  Each transfer moves $4 \cdot 10^8$ elements of the
given data type.}
\label{fig:datatype-speedup}
\end{figure}

The magnitude of the benefit does depend on the types of data
transferred.  \Figref{datatype-speedup} shows the speedup between the
file system and \Name for $4 \cdot 10^8$ elements of various data
types.  As the figure shows, the transfer performance for fixed-width
primitives perform significantly better than string transfers, due
to the smaller amount of data transferred when using
\AugmentedStrings.  While strings do not benefit from our
optimizations, they still benefit from avoiding serializing to the
file system during data transfer.

\subsection{Optimizations}
\label{sec:eval:opt}

We drill down into the different components of the speedup afforded
by \Name's data pipes.  In this section, we evaluate the performance
benefits of \FormOpt's optimizations, which convert text-formatted data to
binary after removing delimiters and metadata.
%exported through a textual format and enable its transmission in a
%binary format without delimiters and metadata.

\subsubsection{String Decorations}

We first evaluate the optimizations due to the use of augmented strings
%implemented in \FormOpt's string
%augmentation mode
as described in \autoref{sec:string-decoration}.
\Figref{manual-vs-automatic} shows the performance of an export
between Myria and Giraph using this mode and individual optimizations.
For this pair, \FormOpt's optimizations are responsible for
approximately one third of the total runtime benefit beyond what
\IORedirect already provides.

To assess the performance benefits of avoiding both text-encoding and
delimiters, we compare the performance against a
manually-constructed data pipe that implements only these two
optimizations. To produce the manually constructed pipes, we modify
each \DBMS to directly transmit/receive results to/from a network
socket and eliminate all intermediate logic related to text-encoding
that might degrade performance.

\input{results/manual-vs-automatic}

In our experiment, we transfer data in both directions and measure
the total runtime. Overall, the \Name-generated data pipes
perform closely to their manually-optimized counterparts. Transferring
from Myria to Giraph is slower due to the implementation of the Giraph
import, where Giraph materializes \AugmentedString instances into character
strings, and processes characters from the materialized string to escape
them if needed.

\subsubsection{Library Extensions}

We generate a library extension implementation for the Jackson JSON
library and evaluate its
performance under the library extension mode of \FormOpt.  Interestingly, for
the pairs of \DBMSs that we examine, most do not support mutually-compatible
exchange using JSON as an intermediate format.  For example, Myria produces a
single JSON document, Spark and Giraph both expect a document-per-line, and
Derby does not natively support bulk import and export of JSON at all.

\Figref{library-extension} shows the performance of using library
extensions with Jackson between Spark and Giraph.  As
before, we use a mutually-compatible JSON adjacency-list format for the
schema of transmitted data.  We find that the relative performance
benefit closely matches that of the string decorations.

\input{results/library-extension}

\subsection{Intermediate Format}
\label{sec:intermediate-format}

Once \FormOpt captures the transferred data in \AugmentedString
%objects before its text encoding, the data pipe type becomes free to
we can use any intermediate format to transfer the data between \DBMSs.  In
this section, we show that, as expected, the choice of that
intermediate format significantly impacts performance, with recent
formats outperforming older ones. This observation is important
as a key contribution of \Name is to free
%to enable the
%evolution of the binary data transfer format and its automatic
%application without requiring
developers from the need to add new data import and
export code to the \DBMS every time a new data format becomes available.

%To
%isolate the performance differences that result from a particular
%choice, we perform a data transfer and measure runtime using several
%format options.

Our experiments include two third-party formats: protocol
buffers~\cite{protobuf} and Arrow~\cite{arrow}.
We also evaluate the basic custom
format from the previous section, which transmits schema information
as a header, fixed width values in binary form, and uses
length-prefixing for strings.  We examine protocol buffers using a
version where message templates are fixed at compile time and another
where they are dynamically constructed at runtime.
\Figref{performance-by-format} shows the results.  Protocol
buffers, depending on whether message formats are statically or
dynamically generated, perform approximately as well as our custom
binary format.  The recent data format, Arrow, offers a
substantial boost in performance, due primarily to its optimized
layout, efficient allocation process, and optimized
iteration~\cite{arrow}.  Additionally, the column-oriented format
offers a further modest advantage over its row-oriented counterpart.

%Given that column-oriented \texttt{ArrowBufs} was the best-performing format,
%our subsequent experiments utilize this format exclusively.

\input{results/intermediate-format}

It is clear that Apache Arrow yields the highest performance as an
intermediate format.  Since we preallocate a buffer for use during
block-oriented transfer necessary for pivoting the data into a
columnar format, it is necessary to reserve an appropriate size for
this intermediate buffer.  In \figref{buffer-size}, we illustrate
transfer performance between  Myria and Giraph for various
\texttt{ArrowBuf} sizes.  Note that since Arrow is column-oriented, we
allocate one buffer for each attribute.  As with our previous
experiments, this includes 4 four-byte integers and 4 eight-byte
doubles.  This implies that for a $n$-element buffer we preallocate
eight \texttt{ArrowBuf}s for a combined total of $48n$ bytes (plus
any bookkeeping overhead required by Arrow's implementation).
As long as the buffer is not too small, the buffer size
has only a negligible impact on performance.

\input{results/buffer-size}

%However, since the size of the data being transferred may easily exceed
%available memory, the data pipe implementation must preallocate a buffer for
%block-oriented transfer.

\subsection{Compression}
\label{sec:compression}

Orthogonal to the choice of data format and application of
optimizations is \Name's ability to compress the data transferred
between pairs of \DBMSs.  Utility of this approach depends on the
distance between \DBMS workers, with nearby \DBMSs being less likely to
benefit than distant ones.

\Figref{compression-transfer} shows the performance of transfer using
compression (or lack thereof) from Myria to Giraph.  We
use three compression techniques: run-length encoding, dictionary-based
compression (zip), and uncompressed transfer.  We separately show transfer
performance between colocated workers (\figref{local-transfer}) and
workers with a 40ms artificial latency introduced into the network adapter
(\figref{remote-transfer}).  For colocated workers, we also evaluate
transfer performance using shared memory; all other transfers take place over
sockets.

For colocated nodes, both compression techniques add modest overhead to
the transfer process that yields a net loss in performance.  For nodes with
higher latency, we show a modest benefit for dictionary-oriented compression.
This suggests that this strategy may be beneficial for physically-separated
\DBMSs.

%While this
%suggests that its application does not improve performance, this may not
%extrapolate to all workloads.  For nodes residing in the same AWS region,
%compression \todo{offered a modest performance advantage.}  This suggests that
%it is advantageous for the worker to dynamically activate compression as a
%function of worker distance rather than as a global option.

\input{results/compression}

\subsection{Code Modifications}
\label{sec:eval:loc}

To get a sense of the amount of changes needed to generate the data pipes
for each \DBMSs,
\tabref{modifications} lists the number of modifications made by the
\IORedirect and \FormOpt components, in terms of the
number of classes and lines of code affected.  We also measure the amount of
time \Name spent to perform the compile-time modifications.

As the results show, the amount of changes is modest across
all of the \DBMSs we evaluate.
In addition, the total number of modifications is small for the \IORedirect
component.  This shows that \DBMS implementations rarely
open an import or export file at more than one call site.  The \FormOpt
component performs more modifications, with library extension requiring fewer
changes than string decoration.  Even with string decoration, for
the \DBMSs that we evaluate, primitive values
are converted to/from a string in close proximity in the code where they are
written/read.  This reduced the number of modifications required for
this optimization.

\input{results/code-modifications}

%% file: results/microbenchmarks.tex
\begin{figure}[t]
\centering
\includegraphics[width=\columnwidth]{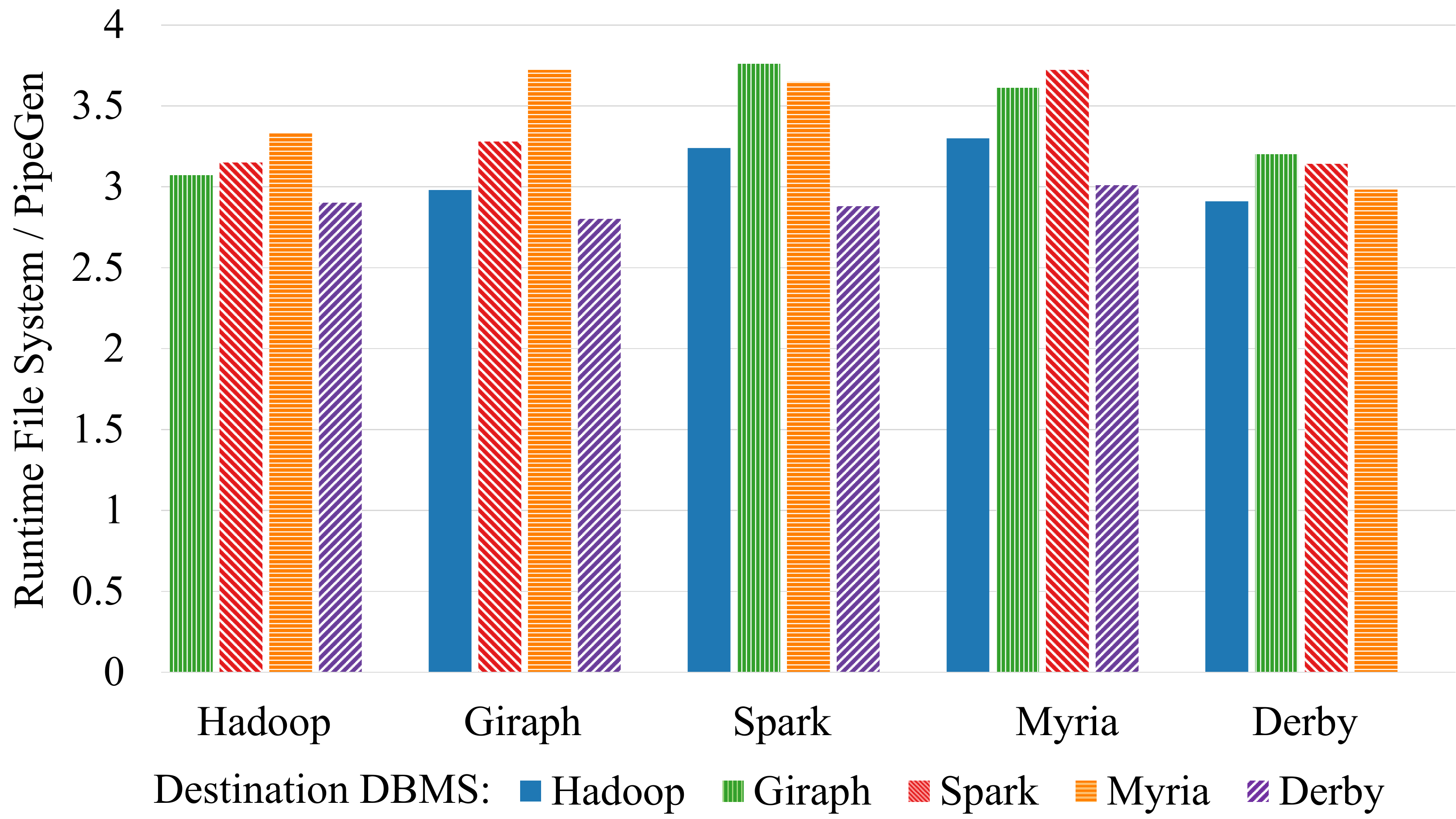}
\vspace{-0.65cm}
\caption{Total speedup between file system and \Name for $10^9$ elements.  Transfer occurred from a source \DBMS ($x$-axis) to a destination \DBMS (bar color/pattern) using the \CSV format.  The number of workers/tasks was fixed at 16.}
\label{fig:microbenchmarks}
\end{figure}

%% file: results/data-types.tex
\begin{table}[t]
\centering
\begin{tabular}{rcccc}
  \toprule
  \# Workers & 1 & 4 & 8 & 16 \\
  \midrule
  Speedup & 3.1 & 3.7 & 3.5 & 3.7 \\
  \bottomrule
\end{tabular}
\vspace{-0.25cm}
\caption{Overall speedup (file system / \Name runtime)
from Myria to Spark for $4 \cdot 10^8$ elements when varying the number
of workers and tasks involved in the transfer.}
\label{t:data-types}
\end{table}

%% file: results/manual-vs-automatic.tex
\begin{figure*}[t]
\centering

\includegraphics[width=\columnwidth, trim=0 2.5cm 0 0, clip=true]{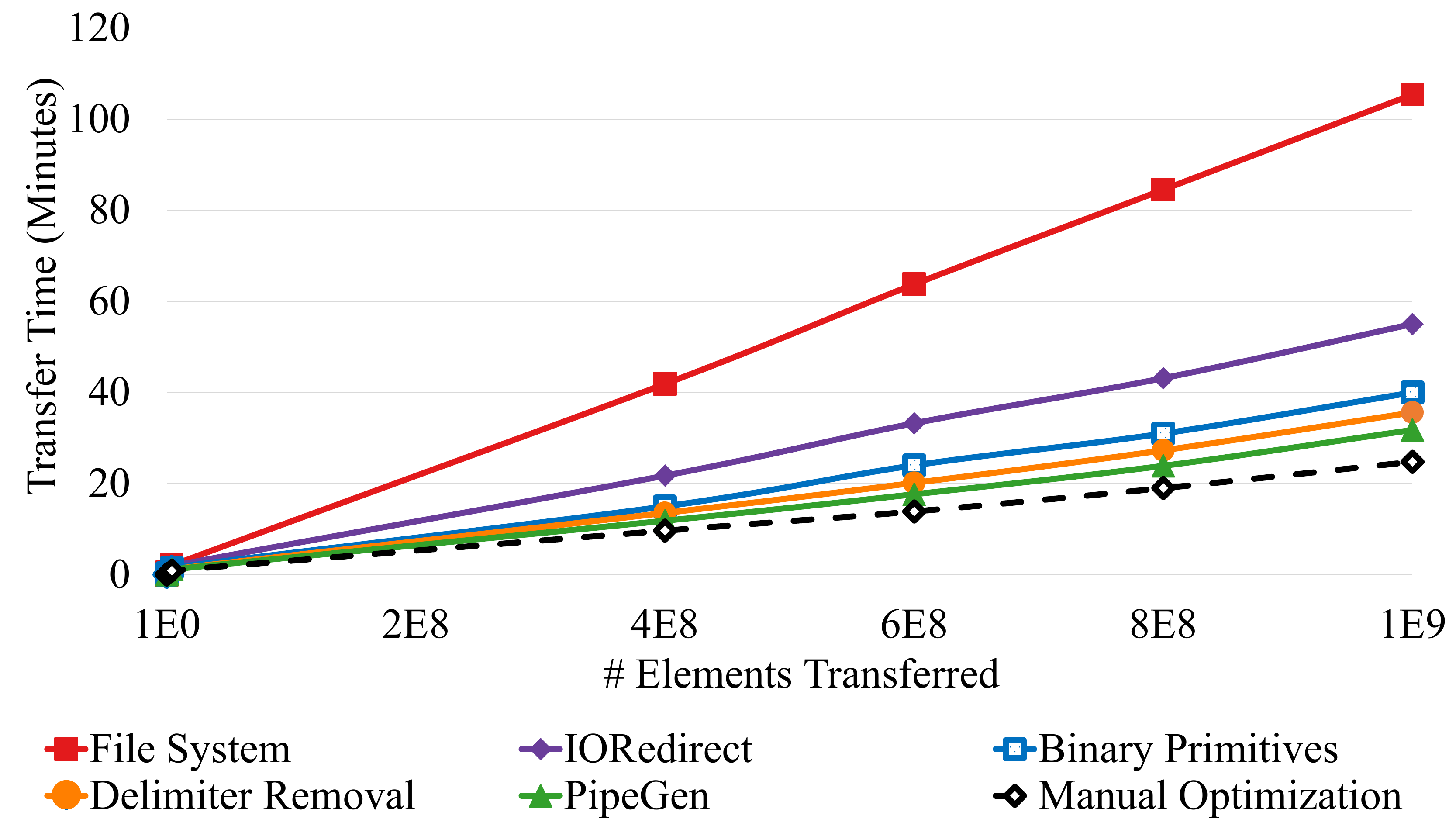}
\includegraphics[width=\columnwidth, trim=0 2.5cm 0 0, clip=true]{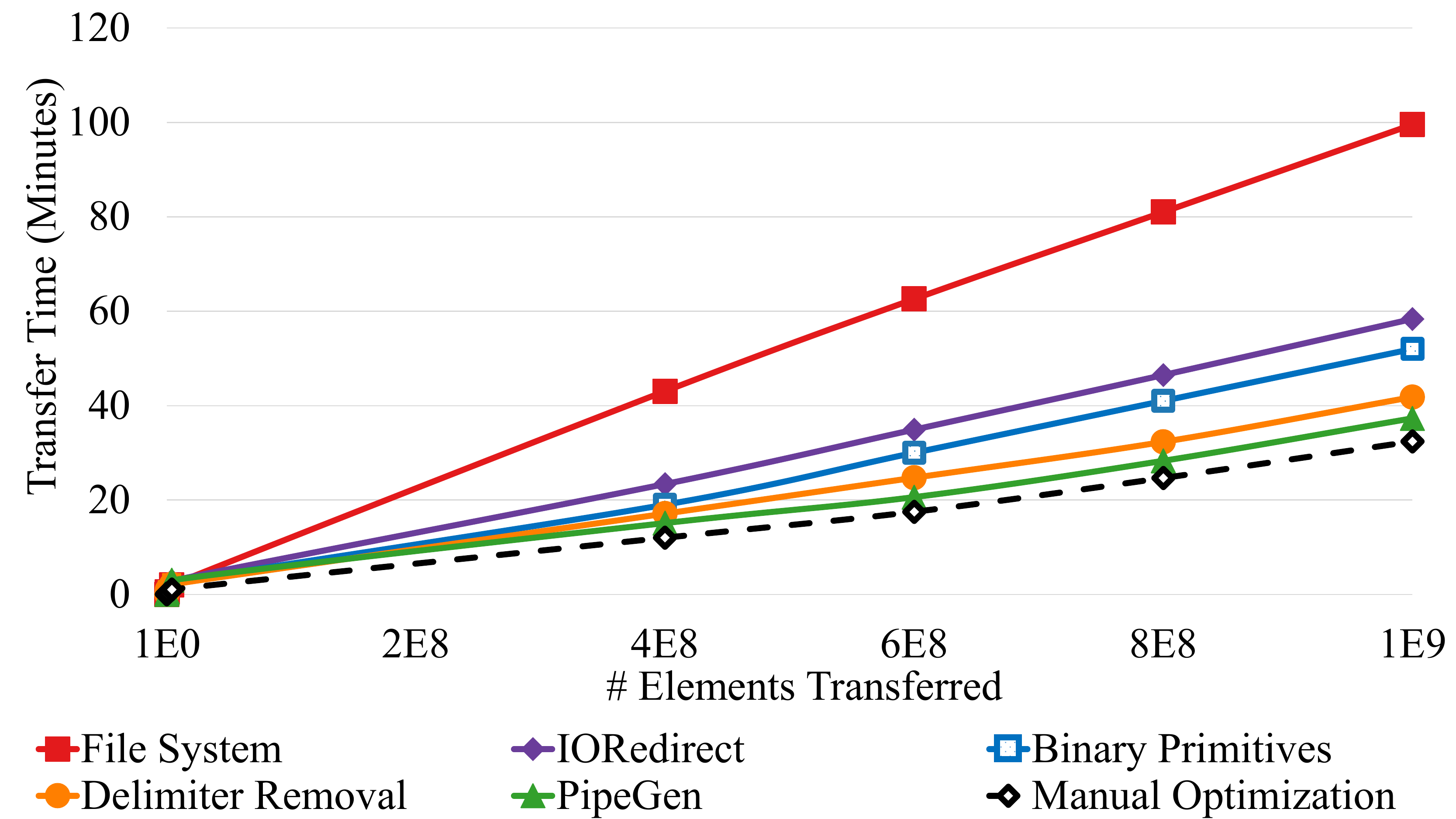}
\includegraphics[width=\columnwidth, trim=0 0 0 17cm, clip=true]{results/manual-comparison1.pdf}

\vspace{-0.3cm}

\subfigure[Giraph$\rightarrow$\Myria]{
  \label{fig:giraph-to-myria}
  \hspace{\columnwidth}
}
\subfigure[\Myria$\rightarrow$Giraph]{
  \label{fig:myria-to-giraph}
  \hspace{\columnwidth}
}
\vspace{-0.3cm}
\caption{Transfer time between the Myria and Giraph using
  \Name and a
  manually-optimized variant.  In (a) we export tuples from Myria and
  import them as vertices in Giraph.  In (b) we reverse the direction
  of transfer.
  We show as baseline transfer through the file system.
  We then activate \Name optimizations as follows.  First, we apply the \IORedirect component.  Next,
  we transmit fixed-width values in binary form.  We then activate
  delimiter removal.  The PipeGen series
  shows all optimizations, which additionally include column pivoting.}
\label{fig:manual-vs-automatic}
\end{figure*}

%% file: results/library-extension.tex
\begin{figure}[t]
\centering
\includegraphics[width=\columnwidth]{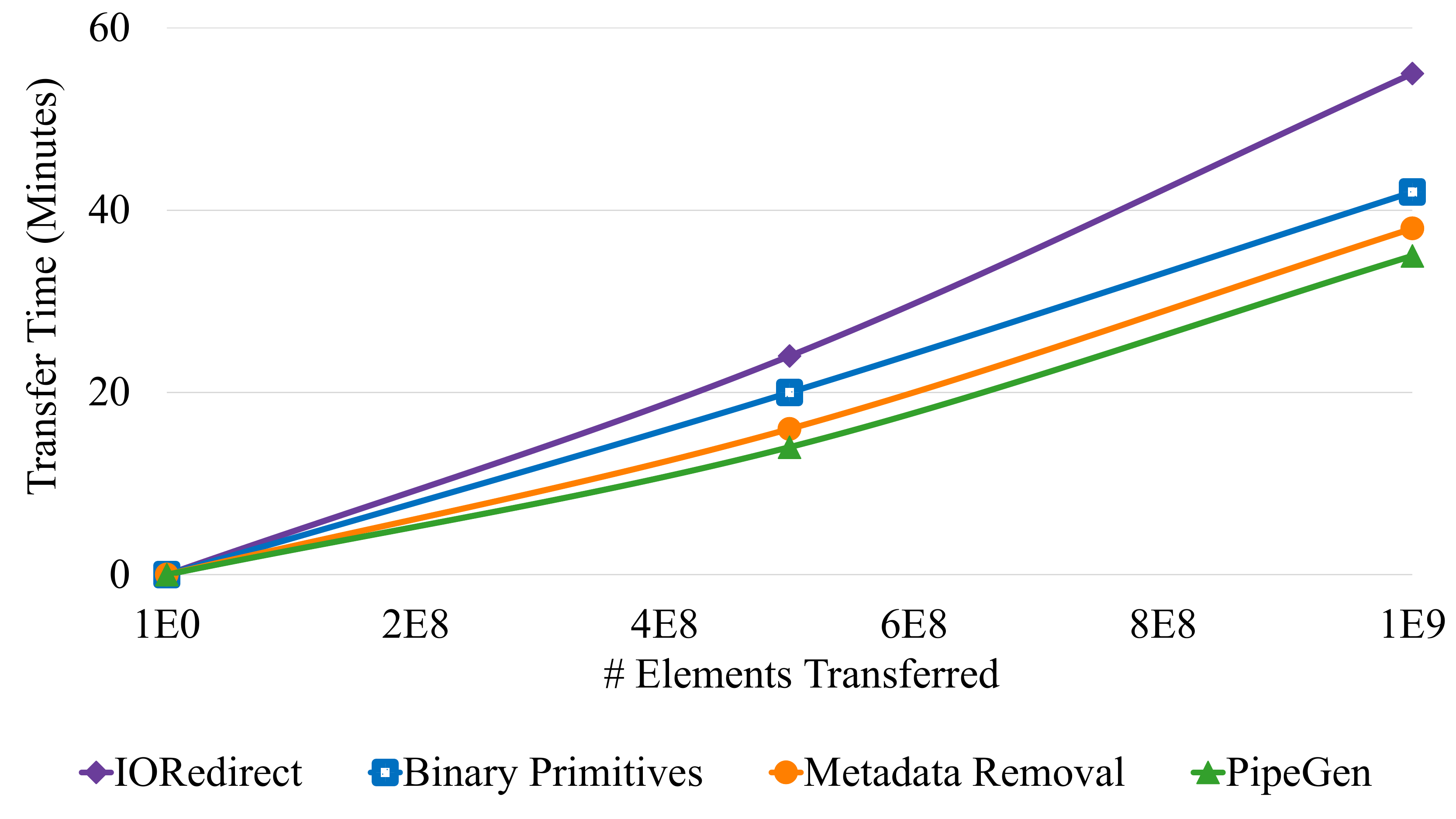}
\vspace{-0.6cm}
\caption{Total runtime of a transfer from Spark to Giraph using the
  library extension mode for the Jackson JSON library.
  We show a baseline application of only the \IORedirect component.  We then
  transmit fixed-width values in binary form.  Next, we activate
  metadata removal (\ie, repeated column names and delimiters).  The PipeGen series
  shows all optimizations (\ie, column pivoting) activated.}
\label{fig:library-extension}
\end{figure}

%% file: results/intermediate-format.tex
\begin{figure}[t]
\centering
\includegraphics[width=\columnwidth]{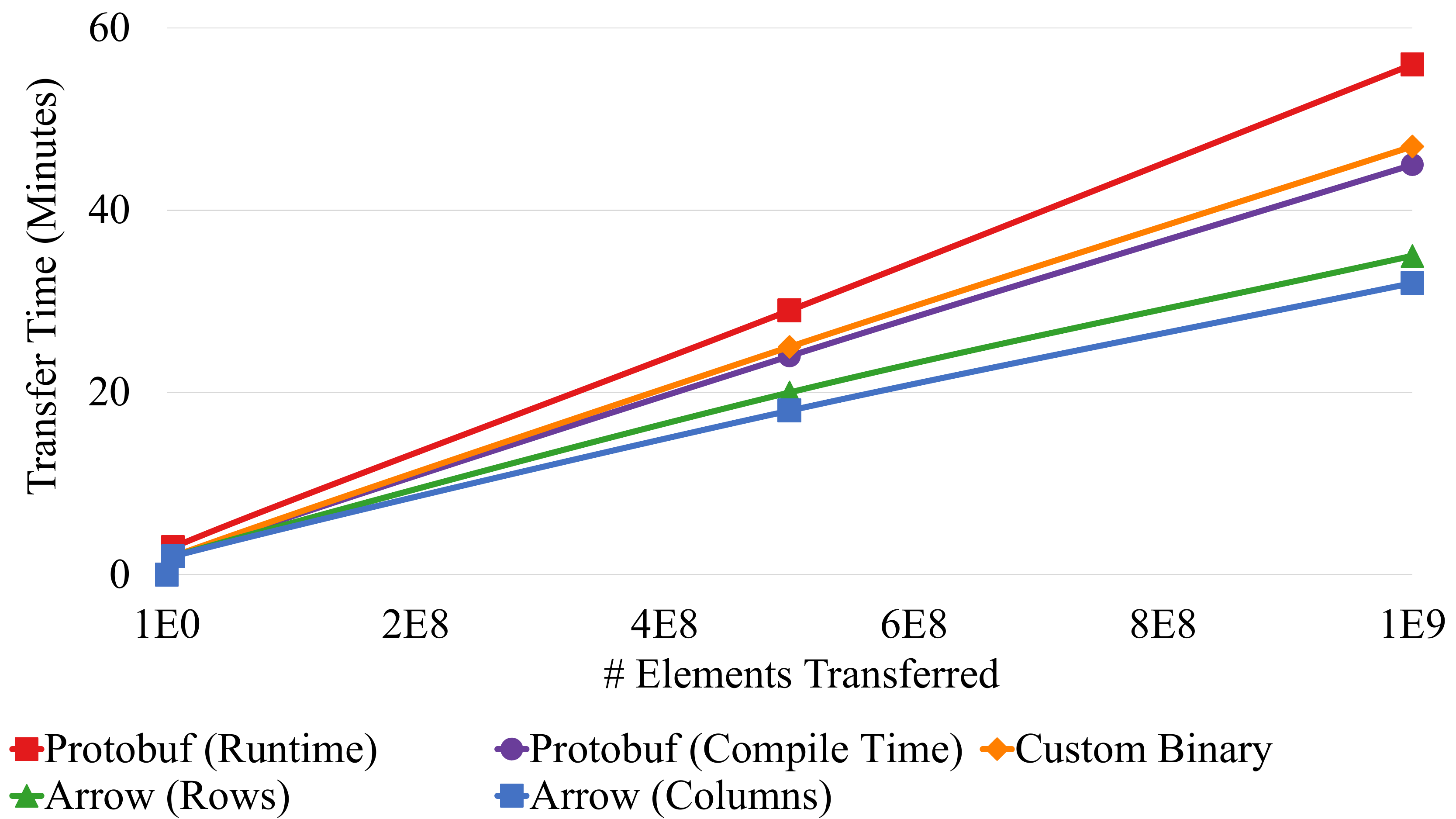}
\vspace{-0.6cm}
\caption{Transfer performance by intermediate format between Hadoop and
Spark.  Message templates for protocol buffers were generated both at
compile time and dynamically at runtime.}
\label{fig:performance-by-format}
\end{figure}

%% file: results/buffer-size.tex
\begin{figure}[t]
\centering
\includegraphics[width=\columnwidth]{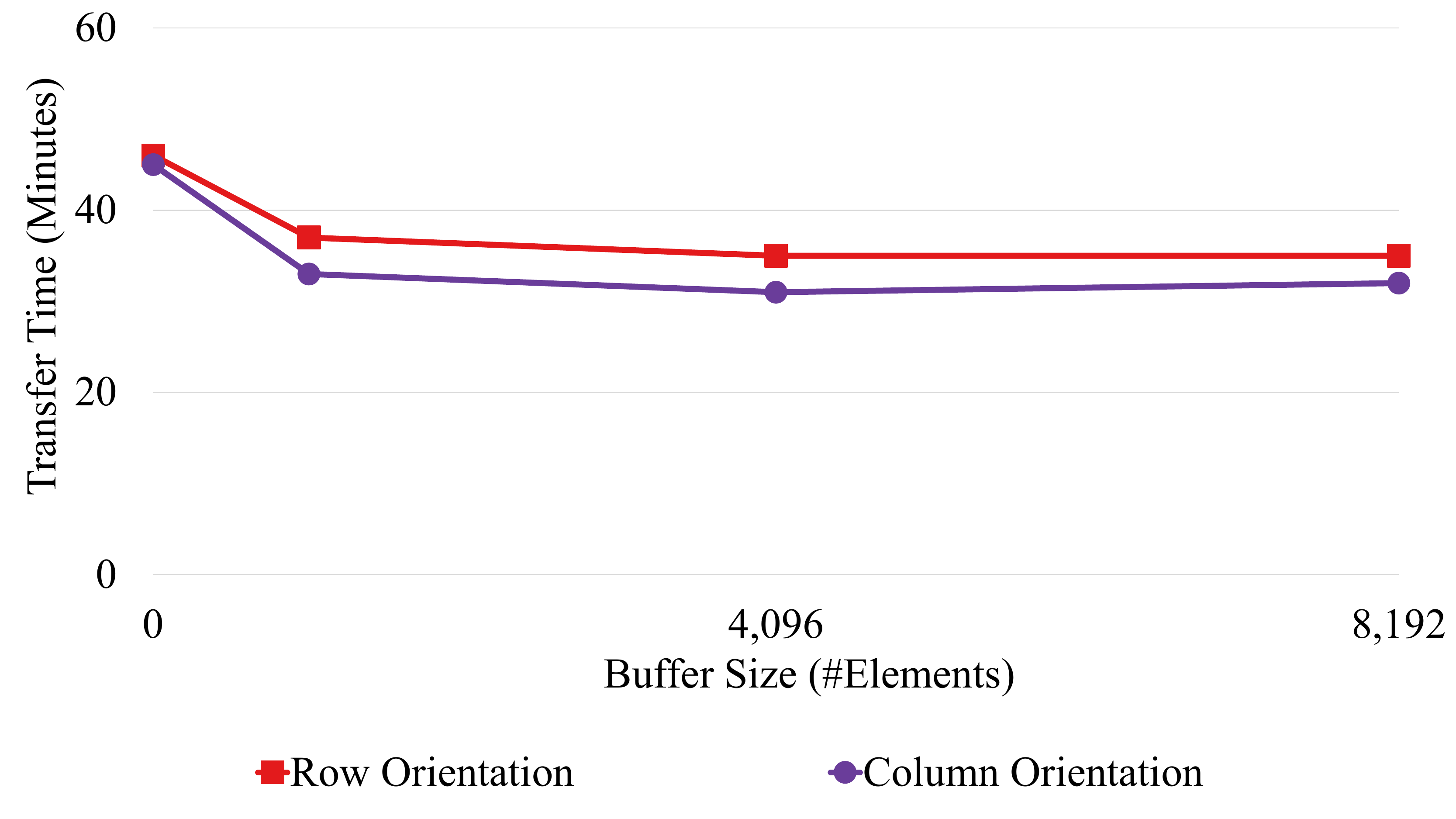}
\vspace{-0.75cm}
\caption{Transfer time from Myria to Giraph for $10^9$ elements with varying sizes of Arrow buffers.
Each of the column buffers used for the transfer was sized to hold the number of values listed on the $x$-axis.
}
\label{fig:buffer-size}
\end{figure}

%% file: results/compression.tex
\begin{figure*}[t]
\centering

\includegraphics[width=\columnwidth, trim=0 2.5cm 0 0, clip=true]{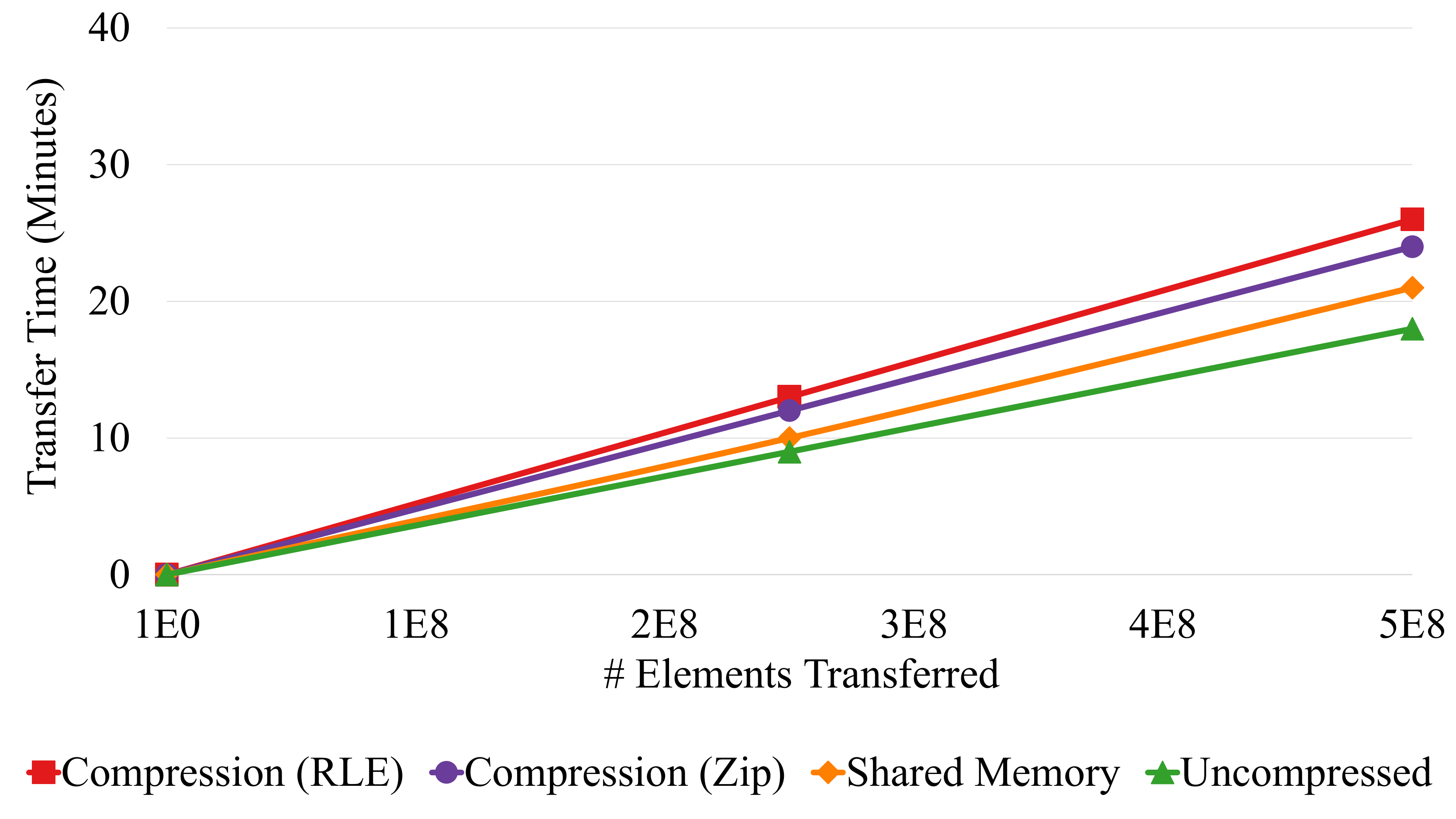}
\includegraphics[width=\columnwidth, trim=0 2.5cm 0 0, clip=true]{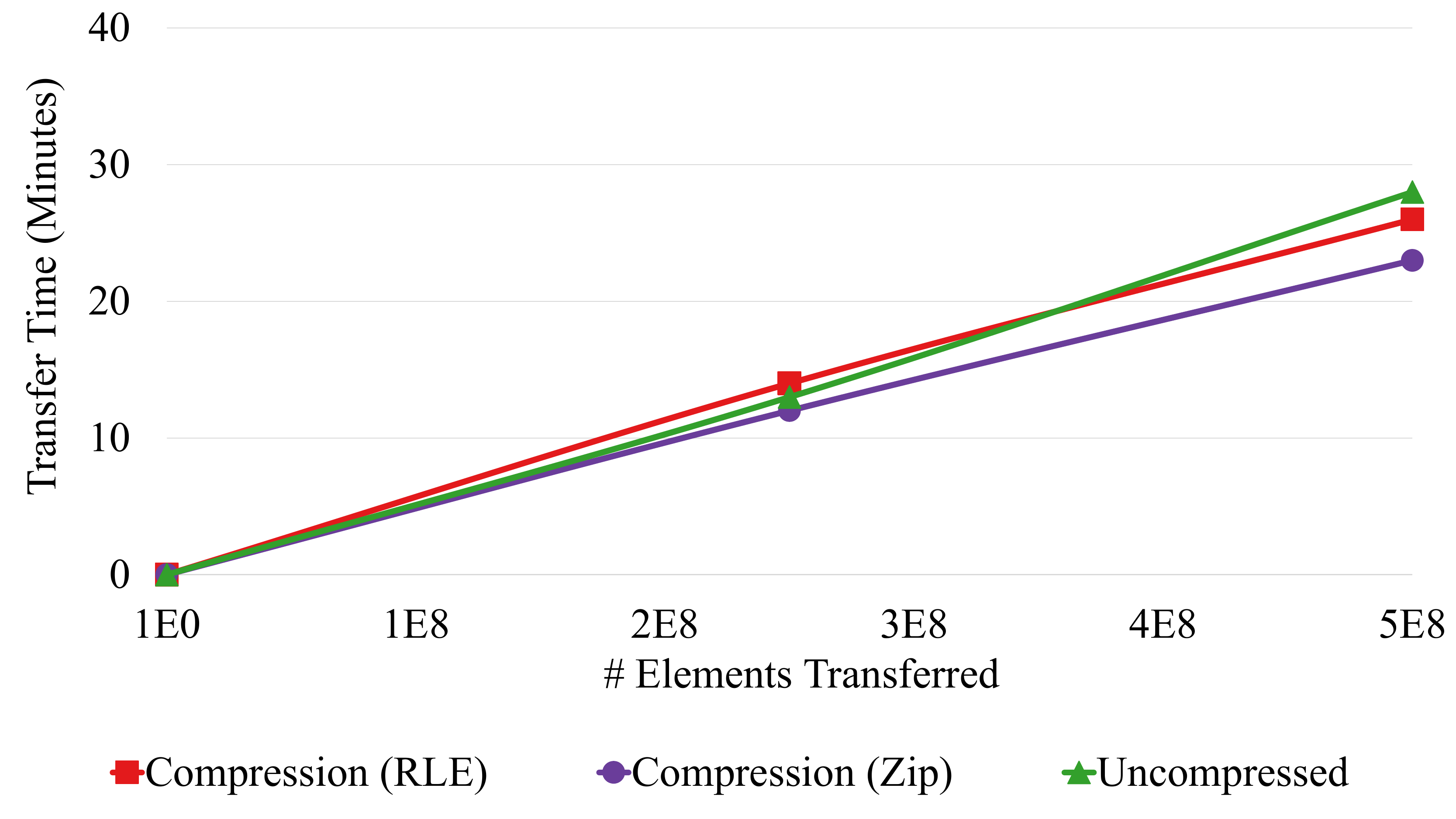}
\includegraphics[width=\columnwidth, trim=0 0 0 17cm, clip=true]{results/local-transfer.pdf}

\vspace{-0.3cm}

\subfigure[Myria$\rightarrow$ Giraph, colocated workers]{
  \label{fig:local-transfer}
    \hspace{\columnwidth}
}
\subfigure[Myria$\rightarrow$ Giraph, 40ms latency]{
  \label{fig:remote-transfer}
    \hspace{\columnwidth}
}

\vspace{-0.4cm}
\caption{Transfer time between Myria and Giraph
using compression.  In (a) we transfer between
workers colocated in the same physical node, while in (b) we transfer
between workers with 40ms latency artificially introduced.
For RLE, zip, and uncompressed formats we transfer
data using a socket.  For colocated nodes, we also transfer data uncompressed
using shared memory.} \label{fig:compression-transfer}
\end{figure*}

%% file: results/code-modifications.tex
\begin{table*}[t]
\centering
\small
\begin{tabular}{ccc@{\qquad}cc@{\qquad}cc}
  \toprule
  \multirow{2}{*}{Mode} &
  \multirow{2}{*}{\DBMS} &
  \multirow{2}{*}{Modification Time} &
  \multicolumn{2}{c}{\IORedirect} &
  \multicolumn{2}{c}{\FormOpt} \\

  \cmidrule{4-7}

  & & (seconds) & \#Classes & LOC &
                  \#Classes & LOC \\
  \midrule
  String Decoration
  & Hadoop       & 245 & 3 &  6 & 6 & 36 \\
  & Myria       & 160 & 2 &  8 & 5 & 54 \\
  & Giraph       & 223 & 2 &  9 & 4 & 47 \\
  & Spark        & 187 & 5 & 18 & 8 & 38 \\
  & Derby        &  130 & 2 &  5 & 2 & 67 \\
  \midrule
  Library Extension
  & Spark        & 178 & 5 & 18 & 2 & 6 \\
%  & Myria        & \todo{0} & \todo{0} & \todo{0} & \todo{0} & \todo{0} \\
  \bottomrule
\end{tabular}
\vspace{-0.2cm}
\caption{For each of the DBMSs we evaluated, we show the time required for
\Name to modify each system.  In addition, for each of the two phases, we also show the
number of classes and lines of code modified by each component.}
\label{t:modifications}
\end{table*}

%% file: related-work.tex
There have been a number of systems
that implement hybrid data analytics using an orchestrator-based
architecture over various constituent \DBMSs.
In such systems, each constituent must be
manually integrated into the hybrid engine.
This architecture was present in
the earliest heterogeneous systems such as COSYS
\cite{adiba1978cooperation}, HD-DBMS~\cite{cardenas1987heterogeneous}, MDAS
\cite{desai1992mdas}, InterViso~\cite{templeton1995interviso}, and Tukwila
\cite{ives1999adaptive}, and is common in more recent systems such as the
Cyclops~\cite{cyclops}, Polystore~\cite{polybase}, Estocada
\cite{bugiotti2015toward}, and Musketeer~\cite{gog2015musketeer}. Our data
pipes complement these systems by providing an efficient mechanism to transfer
large amounts of data between them.
%for which little attention has been paid to this problem in the past.
%automatically exposing what is effectively a
%simple API across each pair of child \DBMSs and enables direct transfer between
%children.

Data exchange is another related topic that also involves transferring of
data across heterogeneous systems.  However, while there has been much work
on generating robust mappings between
schemata across systems~\cite{fagin2005data},
optimization has focused
primarily on inference performance~\cite{saleem2008porsche,haas2005clio},
or does not address data shipping performance~\cite{risch2004functional}.
In contrast, \Name focuses on optimizing
data transfer performance, and assumes that the user or
a query optimizer can generate
query plans to reconcile the different schemata across \DBMSs.

\Name automatically embeds data pipes into a selected \DBMS.  Some previous
work in the area of data integration has explored the manual embedding of
data pipes for transferring data across heterogeneous systems.
For example, Rusinkiewicz et al.
proposed a common inter-\DBMS format mediated via STUB operators
\cite{rusinkiewicz1988distributed}.  These operators are similar to data pipes,
but are manually generated and do not address direct transfer between \DBMSs.
Additionally, prior work has explored the automatic generation of data
transformation operators that are
similar to data pipes.  For example, Mendell et al. used code generation
under a general-purpose streaming engine to support streaming XML processing
\cite{mendell2012extending}.  Similar tools target automatic parsing of
semi-structured data for ad hoc processing~\cite{fisher2008dirt}.  While these
efforts shares some commonality with \Name, they target a single data model and
do not address data transfer performance between \DBMSs.

There are many potential formats
(\eg, protocol buffers~\cite{protobuf}, Parquet~\cite{parquet},
Thrift~\cite{thrift}, Avro~\cite{avro}, Arrow~\cite{arrow}) that can be used
to move data between two \DBMSs.  Since each of these requires manual effort
in order to be supported by a \DBMS, \Name's ability to switch between formats
as they evolve allows users to capture performance benefits with reduced effort.

Finally, \Name's \IORedirect component redirects specific file
system calls to network socket operations instead.  This
redirection is similar to previous work in the systems research community.
For example, the
CDE software distribution tool intercepts file system calls using
\texttt{ptrace} and redirects \IO to an alternate location \cite{guo2011cde};
similar work by Spillane et al. used system call redirection for rapid
prototyping~\cite{spillane2007rapid}.  This approach is also common in the
security community for applications ranging from sandboxing~\cite{wagner1996secure}
to information flow analysis~\cite{zeldovich2006making}.

%% file: conclusion.tex
In this paper we described \Name, a tool that automatically generates data
pipes between \DBMSs for efficient data transfer.
Observing that most DBMSs support importing from and exporting to
a common data format, \Name makes use of
existing test cases and program analysis techniques
to create data pipes from existing source
code, and furthermore optimizes the generated implementation by
removing several inefficiencies in data encoding and unnecessary
data serialization.

%inefficiencies in text oriented formats and implemented a prototype
%demonstrating performance gain for hybrid OLAP analytics.
We have implemented a prototype of \Name and evaluated it by automatically generating
data pipes across a variety of \DBMSs, including relational and
graph-based engines. Our experiments show that the \Name-generated
data pipes enables efficient hybrid data analytics by
outperforming the traditional way of transferring data
via the file system by up to $3.8\times$.
%As future work, we plan to target
%other common formats (e.g., Hadoop sequence files).
%and evaluate whether the
%optimization of binary formats (e.g., via shared memory access) are viable.
%In addition, we will integrate \Name into an orchestrator and measure
%end-to-end performance improvement in executing data analytic queries.

%% file: automatic-connectors.bbl
\begin{thebibliography}{10}

\bibitem{adiba1978cooperation}
M.~Adiba and D.~Portal.
\newblock A cooperation system for heterogeneous data base management systems.
\newblock {\em Information Systems}, 3(3):209--215, 1978.

\bibitem{JDBC}
L.~Andersen.
\newblock Jdbc 4.2.
\newblock Technical Report JSR 221, Oracle, March 2014.

\bibitem{apache-commons-csv}
{Apache Commons CSV}.
\newblock \url{https://commons.apache.org/proper/commons-csv/}.

\bibitem{derby}
{Apache Software Foundation}.
\newblock Derby.
\newblock \url{https://db.apache.org/derby}.

\bibitem{hadoop}
{Apache Software Foundation}.
\newblock Hadoop.
\newblock \url{https://hadoop.apache.org}.

\bibitem{arrow}
Apache arrow.
\newblock \url{https://arrow.apache.org/}.

\bibitem{avery2011giraph}
C.~Avery.
\newblock Giraph: Large-scale graph processing infrastructure on hadoop.
\newblock {\em Proceedings of the Hadoop Summit. Santa Clara}, 2011.

\bibitem{avro}
{Apache Avro}.
\newblock \url{https://avro.apache.org/}.

\bibitem{bugiotti2015toward}
F.~Bugiotti, D.~Bursztyn, A.~Deutsch, I.~Ileana, and I.~Manolescu.
\newblock Toward scalable hybrid stores.
\newblock In {\em SEBD}, 2015.

\bibitem{cardenas1987heterogeneous}
A.~F. Cardenas.
\newblock Heterogeneous distributed database management: The hd-dbms.
\newblock {\em Proceedings of the IEEE}, 75(5):588--600, 1987.

\bibitem{datoblog}
{Dato, Inc.}
\newblock Graphlab integration with spark open source release.
\newblock \url{http://blog.dato.com/graphlab-integration-with-spark}, 2015.

\bibitem{desai1992mdas}
B.~C. Desai and R.~Pollock.
\newblock Mdas: heterogeneous distributed database management system.
\newblock {\em Information and Software Technology}, 34(1):28--42, 1992.

\bibitem{polybase}
D.~J. DeWitt, A.~Halverson, R.~Nehme, S.~Shankar, J.~Aguilar-Saborit,
  A.~Avanes, M.~Flasza, and J.~Gramling.
\newblock Split query processing in polybase.
\newblock In {\em SIGMOD}, pages 1255--1266. ACM, 2013.

\bibitem{elmore2015demonstration}
A.~Elmore, J.~Duggan, M.~Stonebraker, M.~Balazinska, U.~Cetintemel,
  V.~Gadepally, J.~Heer, B.~Howe, J.~Kepner, T.~Kraska, et~al.
\newblock A demonstration of the bigdawg polystore system.
\newblock {\em VLDB}, 8(12):1908--1911, 2015.

\bibitem{fagin2005data}
R.~Fagin, P.~G. Kolaitis, R.~J. Miller, and L.~Popa.
\newblock Data exchange: semantics and query answering.
\newblock {\em Theoretical Computer Science}, 336(1):89--124, 2005.

\bibitem{fisher2008dirt}
K.~Fisher, D.~Walker, K.~Q. Zhu, and P.~White.
\newblock From dirt to shovels: Fully automatic tool generation from ad hoc
  data.
\newblock In {\em POPL}, page 421. ACM, 2008.

\bibitem{gog2015musketeer}
I.~Gog, M.~Schwarzkopf, N.~Crooks, M.~P. Grosvenor, A.~Clement, and S.~Hand.
\newblock Musketeer: all for one, one for all in data processing systems.
\newblock In {\em Proceedings of the Tenth European Conference on Computer
  Systems}, page~2. ACM, 2015.

\bibitem{guo2011cde}
P.~J. Guo and D.~R. Engler.
\newblock Cde: Using system call interposition to automatically create portable
  software packages.
\newblock In {\em USENIX ATC}, 2011.

\bibitem{haas2005clio}
L.~M. Haas, M.~A. Hern{\'a}ndez, H.~Ho, L.~Popa, and M.~Roth.
\newblock Clio grows up: from research prototype to industrial tool.
\newblock In {\em SIGMOD}, page 805. ACM, 2005.

\bibitem{myria}
D.~Halperin, V.~T. de~Almeida, L.~L. Choo, S.~Chu, P.~Koutris, D.~Moritz,
  J.~Ortiz, V.~Ruamviboonsuk, J.~Wang, A.~Whitaker, et~al.
\newblock Demonstration of the myria big data management service.
\newblock In {\em SIGMOD}, pages 881--884. ACM, 2014.

\bibitem{ives1999adaptive}
Z.~G. Ives, D.~Florescu, M.~Friedman, A.~Levy, and D.~S. Weld.
\newblock An adaptive query execution system for data integration.
\newblock In {\em ACM SIGMOD Record}, volume~28, pages 299--310. ACM, 1999.

\bibitem{jackson}
{Jackson JSON Processor}.
\newblock \url{http://wiki.fasterxml.com/JacksonHome/}.

\bibitem{jetley2008massively}
P.~Jetley, F.~Gioachin, C.~Mendes, L.~V. Kale, and T.~Quinn.
\newblock Massively parallel cosmological simulations with changa.
\newblock In {\em IPDPS}, pages 1--12. IEEE, 2008.

\bibitem{josifovski2002garlic}
V.~Josifovski, P.~Schwarz, L.~Haas, and E.~Lin.
\newblock Garlic: a new flavor of federated query processing for db2.
\newblock In {\em SIGMOD}, pages 524--532. ACM, 2002.

\bibitem{knebe2013structure}
A.~Knebe, F.~R. Pearce, H.~Lux, Y.~Ascasibar, P.~Behroozi, J.~Casado, C.~C.
  Moran, J.~Diemand, K.~Dolag, R.~Dominguez-Tenreiro, et~al.
\newblock Structure finding in cosmological simulations: the state of affairs.
\newblock {\em MNRAS}, 435(2):1618, 2013.

\bibitem{cyclops}
H.~Lim, Y.~Han, and S.~Babu.
\newblock How to fit when no one size fits.
\newblock In {\em CIDR}, volume~4, page~35, 2013.

\bibitem{lin2010power}
F.~Lin and W.~W. Cohen.
\newblock Power iteration clustering.
\newblock In {\em ICML}, page 655, 2010.

\bibitem{mendell2012extending}
M.~Mendell, H.~Nasgaard, E.~Bouillet, M.~Hirzel, and B.~Gedik.
\newblock Extending a general-purpose streaming system for xml.
\newblock In {\em EDBT}, page 534. ACM, 2012.

\bibitem{myriawebsite}
Myria: {Big Data} management as a {Cloud} service.
\newblock \url{http://myria.cs.washington.edu/}.

\bibitem{ODBC}
{Microsoft Open Database Connectivity (ODBC)}.
\newblock \url{https://msdn.microsoft.com/en-us/library/ms710252}.

\bibitem{parquet}
{Apache Parquet}.
\newblock \url{https://parquet.apache.org/}.

\bibitem{protobuf}
{Protocol Buffers}.
\newblock \url{https://developers.google.com/protocol-buffers/}.

\bibitem{risch2004functional}
T.~Risch, V.~Josifovski, and T.~Katchaounov.
\newblock Functional data integration in a distributed mediator system.
\newblock In {\em The Functional Approach to Data Management}, pages 211--238.
  Springer, 2004.

\bibitem{rusinkiewicz1988distributed}
M.~Rusinkiewicz, K.~Loa, and A.~K. Elmagarmid.
\newblock Distributed operation language for specification and processing of
  multidatabase applications.
\newblock 1988.

\bibitem{saleem2008porsche}
K.~Saleem, Z.~Bellahsene, and E.~Hunt.
\newblock Porsche: Performance oriented schema mediation.
\newblock {\em Information Systems}, 33(7):637--657, 2008.

\bibitem{spillane2007rapid}
R.~P. Spillane, C.~P. Wright, G.~Sivathanu, and E.~Zadok.
\newblock Rapid file system development using ptrace.
\newblock In {\em ExpCS}, page~22. ACM, 2007.

\bibitem{stonebraker:15}
M.~Stonebraker.
\newblock {ACM SIGMOD} blog: The case for polystores.
\newblock \url{http://wp.sigmod.org/?p=1629}.

\bibitem{stonebraker2005c}
M.~Stonebraker, D.~J. Abadi, A.~Batkin, X.~Chen, M.~Cherniack, M.~Ferreira,
  E.~Lau, A.~Lin, S.~Madden, E.~O'Neil, et~al.
\newblock C-store: a column-oriented dbms.
\newblock In {\em VLDB}, pages 553--564. VLDB Endowment, 2005.

\bibitem{scidb}
M.~Stonebraker, P.~Brown, A.~Poliakov, and S.~Raman.
\newblock The architecture of scidb.
\newblock In {\em SSDBM}, pages 1--16. Springer, 2011.

\bibitem{templeton1995interviso}
M.~Templeton, H.~Henley, E.~Maros, and D.~J. Van~Buer.
\newblock Interviso: Dealing with the complexity of federated database access.
\newblock {\em VLDB}, 4(2):287--318, 1995.

\bibitem{thrift}
{Apache Thrift}.
\newblock \url{https://thrift.apache.org/}.

\bibitem{yarn}
V.~K. Vavilapalli, A.~C. Murthy, C.~Douglas, S.~Agarwal, M.~Konar, R.~Evans,
  T.~Graves, J.~Lowe, H.~Shah, S.~Seth, et~al.
\newblock Apache hadoop yarn: Yet another resource negotiator.
\newblock In {\em SOCC}, page~5. ACM, 2013.

\bibitem{wagner1996secure}
D.~Wagner, I.~Goldberg, and R.~Thomas.
\newblock A secure environment for untrusted helper applications.
\newblock In {\em Proc. of the 6th USENIX Unix Security Symp}, 1996.

\bibitem{spark}
M.~Zaharia, M.~Chowdhury, M.~J. Franklin, S.~Shenker, and I.~Stoica.
\newblock Spark: cluster computing with working sets.
\newblock In {\em HotCloud}, pages 10--10, 2010.

\bibitem{zeldovich2006making}
N.~Zeldovich, S.~Boyd-Wickizer, E.~Kohler, and D.~Mazi{\`e}res.
\newblock Making information flow explicit in histar.
\newblock In {\em OSDI}, pages 263--278. USENIX Association, 2006.

\end{thebibliography}
